\documentclass{aa}  

\usepackage[switch]{lineno}
\usepackage{graphicx}
\usepackage{comment}
\usepackage[utf8]{inputenc}
\DeclareUnicodeCharacter{2212}{-}
\usepackage{txfonts}
\usepackage{xcolor}
\usepackage{comment}
\usepackage[normalem]{ulem}
\usepackage{longtable}
\usepackage{appendix}
\usepackage{orcidlink}
\usepackage{amsmath}
\usepackage{comment}
\usepackage{hyperref}

\usepackage[normalem]{ulem}

\begin{document} 

\title{Properties of Chromium and Manganese in Tycho's SNR with XRISM/Resolve}

   \author{R. Giuffrida\orcidlink{0000-0002-2774-3491}\inst{1}, A. Decourchelle \orcidlink{0000-0002-1796-758X}\inst{1}, L. Godinaud \orcidlink{0009-0005-0005-0272}\inst{1}, T. Tanaka \orcidlink{0000-0002-4383-0368}\inst{2}, G. Ferrand \orcidlink{0000-0002-4231-8717}\inst{3,4}, J. Ballet \orcidlink{0000-0002-8784-2977}\inst{1}, T. Holland-Ashford \orcidlink{0000-0002-7643-0504}\inst{5}, C. Badenes \orcidlink{0000-0003-3494-343X}\inst{6}, A. Bamba \orcidlink{0000-0003-0890-4920} \inst{7}, B. J. Williams \orcidlink{0000-0003-2063-381X}\inst{5}, T. Mizuno \orcidlink{0000-0001-7263-0296}\inst{8}, F. Acero \orcidlink{0000-0002-6606-2816}\inst{1,9}}

\institute{Université Paris-Saclay and Université Paris Cité, CEA, CNRS, AIM, F-91191 Gif sur Yvette, France
\and
Department of Physics, Konan University, Hyogo 658-8501, Japan
\and
Department of Physics and Astronomy, 30A Sifton Road, University of Manitoba, Winnipeg, Manitoba, R3T 2N2, Canada
\and
RIKEN Center for Interdisciplinary Theoretical and Mathematical Sciences (iTHEMS), 2-1 Hirosawa, Wako, Saitama 351-0198, Japan
\and
Astrophysics Science Division, NASA Goddard Space Flight Center, Greenbelt, MD 20771, USA
\and
Department of Physics and Astronomy and Pittsburgh Particle Physics, Astrophysics and Cosmology Center (PITT PACC), University of Pittsburgh, 3941 O’Hara Street, Pittsburgh, PA 15260, USA
\and
Department of Physics, University of Tokyo, Tokyo 113-0033, Japan
\and
Hiroshima Astrophysical Science Center, Hiroshima University, Hiroshima 739-8526, Japan
\and
FSLAC IRL 2009, CNRS/IAC, La Laguna, Tenerife, Spain
}

\abstract
   {The X-ray emission of young supernova remnants gives access to the composition and spatial distribution of elements ejected in supernovae. The emission from the ejecta provides crucial insights into the explosion mechanism, especially for Fe-group elements formed in the innermost layers of the star.
   }
   {The objective of this paper is to provide constraints on the supernova progenitor and explosion mechanism of Tycho's type Ia supernova remnant by studying the properties of Cr and Mn.}
   {We analyzed \textit{XRISM}/Resolve high-resolution X-ray spectroscopy observations of five regions of Tychos's supernova remnant, performed during the Performance Verification phase and the AO1 phase. Gaussian line modelling was first used to determine the spatial distribution and intensity of Cr and Mn K$\alpha$ lines. We then modeled the emission of the Cr/Mn/Fe plasma in each pointing with two non equilibrium ionization shock models to reproduce both the front and rear emitting sides of the shocked ejecta shell. Mass ratios of Cr/Fe, Mn/Fe and Mn/Cr are compared with expectations from different type Ia supernova models from 1D and 3D numerical simulations.}
{The unprecedented spectral resolution of \textit{XRISM}/Resolve allows us to detect significant emission lines of Cr and Mn in different regions of Tycho's supernova remnant, even at small scale, providing a much more precise spatial distribution of Cr and Mn with respect to previous measurements. Significant Cr and Mn emission lines are observed in the Fe bright northeastern-to-southeastern emitting rim, some in the center region, but none in the southern region. The observed mass ratios of Mn/Fe, Cr/Fe and Mn/Cr, are consistent with 1D delayed detonation models of near-Chandrasekhar-mass Type Ia supernovae and 1D pure-detonation models of 1 M$_\odot$ sub-Chandrasekhar-mass progenitors. In both cases, the observations imply supersolar zero-age main-sequence progenitor metallicities. 
}
{
Overall the best agreement is found for models with a too high metallicity to be realistic for Tycho's location. This points to additional sources of neutronisation such as carbon simmering for Chandrasekhar mass progenitor. For sub-Chandrasekhar mass, double detonation models with a WD mass lower than 1M$_\odot$ are suggested.}


   \titlerunning{}
   \authorrunning{Giuffrida et al.}

   \maketitle

\section{Introduction} 
\label{sec:intro}

Supernova Remnants (SNRs) are interesting astrophysical sources because they link to the study of chemical evolution of galaxies. The X-ray study of young SNRs allows us to measure the composition and the distribution of elements synthesized during the explosion and serves as a powerful diagnostic tool for examining the supernova (SN) explosion. 
In the case of type Ia SNe, two different scenarios are invoked for the progenitor system: single-degenerate \citep[SD,][]{1973ApJ...186.1007W} and double-degenerate \citep[DD,][]{1984ApJ...277..355W} scenario. 
Both cases involve the thermonuclear explosion of a white dwarf (WD) accreting material from a companion. For the SD scenario, the companion is a non-degenerate star, whether for the DD scenario the companion is another WD (see \citealt{2025A&ARv..33....1R} for an extensive review on the progenitor systems).

One strategy to determine the progenitor system is to focus on the nature of the circumstellar material \citep{2007Sci...317..924P,2011Sci...333..856S,2007ApJ...662..472B,2011ApJ...741...96W,2014ApJ...790..139W} or look for a surviving companion star \citep{2011Natur.480..348L,2012Natur.489..533G,2012Natur.481..164S,2018MNRAS.479.5696K}.
The lack of an observed companion star in the known type Ia SNRs makes the DD scenario the favored one. In this case, a sub-Chandrasekhar mass explosion with low central density is preferred to a near-Chandrasekhar mass explosion with high central density because such a WD would collapse in a neutron star instead of exploding as a type Ia SN \citep{1991ApJ...367L..19N}. 

Different explosion mechanisms have been proposed (see \citealt{2020IAUS..357....1R,2025A&ARv..33....1R} for reviews, and references therein). The explosion is widely believed to occur in a carbon–oxygen white dwarf, either through a delayed detonation scenario in which a near-Chandrasekhar mass white dwarf undergoes a deflagration to detonation transition \citep{1991A&A...245L..25K}, or via a prompt detonation, such as in the double detonation scenario, taking place in a sub-Chandrasekhar mass WD \citep{1994ApJ...423..371W,2010ApJ...714L..52S,2011ApJ...734...38W}. 

A more complex model called “dynamically-driven double-degenerate double-detonation” \citep[D$^6$,][]{2010ApJ...709L..64G,2013ApJ...770L...8P} has recently become promising for type Ia SNe. It naturally produces hypervelocity runaway WDs, as detected with \textit{Gaia} observations, from the ejected companion of the WD that exploded as a type Ia SN \citep{2018ApJ...865...15S}.

The main difference between near-Chandrasekhar mass and sub-Chandrasekhar mass exploding WD for the nucleosynthesis yields arises from the central density of the WD at the onset of the thermonuclear runaway. The iron group elements are very sensitive to these conditions. In particular, Cr, Mn and Ni provide diagnostics on the level of neutronisation in the ejecta. In the dense core ($\rho > 2 \times 10^8$ g cm$^{-3}$) of near-M$_{Ch}$ exploding WDs, the electron capture is favored and produces neutron-rich isotopes $^{54}$Cr, $^{55}$Mn, and $^{58}$Ni \citep{1999ApJS..125..439I}. In deflagration models (eg., \citealt{2014MNRAS.438.1762F}), $^{54}$Cr is the dominant isotope of Cr. However, in most explosion models, Cr is dominated by $^{52}$Cr, which is produced by incomplete Si burning further out and thus probes complementary conditions of the explosion \citep{2013A&A...550A..24B}. Other factors can enter for the Mn production in neutron-rich gas near the center of the explosion, such as the zero-age main sequence metallicity via $^{22}$Ne, the explosion mechanism and the way freeze-out proceeds (normal versus alpha-rich freeze-out), which depends on the central density \citep{2013A&A...559L...5S}. Thus, the Mn/Cr abundance ratio provide constraints on the highest density material in the explosion or the progenitor metallicity. This makes Cr, Mn and Ni powerful diagnostics to study the progenitor and explosion mechanism of type Ia SNe. 

Tycho's SNR is the result a young Galactic type Ia SN, which exploded in 1572 \citep{2006ApJ...645.1373B,2008Natur.456..617K}, located at a distance of 2.3-4 kpc \citep{1980ApJ...235..186C,1994A&A...283..635F,1995A&A...299..193S,2010ApJ...725..894H,2013ApJ...770..129W}. Recent results using light echoes have evaluated the distance for Tycho's SNR to $3.2^{+0.1}_{-0.2}$ kpc \citep{2024OJAp....7E..98N}. Tycho's SNR has been observed with different X-ray telescopes. First of all, \emph{ASCA} observations confirmed the origin as a type Ia SN and showed the presence of emission lines from Si, S and Fe \citep{1997ApJ...475..665H}. Moreover, \emph{XMM-Newton} observations had shown a stratification of the elements, where the Fe is located in the inner regions of the remnant and the lighter elements in the outer parts, in agreement with nucleosynthesis models for type Ia SNe \citep{2001A&A...365L.218D}. The excellent spatial resolution of \emph{Chandra} traced the detailed distribution of the elements and the forward shock \citep{2005ApJ...634..376W}, showing that Rayleigh-Taylor instabilities strongly affect the morphology of the remnant at the contact discontinuity between ejected and ambient material. It has been also shown that in Tycho's SNR the thermal X-ray emission is dominated by the SNR ejecta \citep{2007ApJ...665..315C}, while the non-thermal X-ray emission is mostly due to synchrotron emission at the forward shock \citep{2005ApJ...621..793B,2015ApJ...814..132L}. Some non-thermal X-ray features called "stripes" have also been found in the west and in the south \citep{2011ApJ...728L..28E}.

Tycho's SNR is also a very good laboratory to study the dynamics of ejecta. Early expansion and Doppler measurements were consistent with spherical expansion \citep{2010ApJ...725..894H,2017ApJ...842...28W}.
\cite{2023A&A...680A..80G,2025A&A...693A.234G} have shown, from \textit{Chandra} data, a strong asymmetry  in the Si-rich ejecta mean line-of-sight velocity between the north and the south of the remnant, where the northern part is mostly blueshifted and the southern part is mostly redshifted. With \textit{XRISM}/Resolve, a front/rear velocity asymmetry of the ejecta of Tycho’s SNR is revealed (XRISM collaboration, in prep , \cite{2026A&Asubn}).
Regarding the explosion mechanism, \cite{2006ApJ...645.1373B} have shown that models with a delayed detonation transition (DDT) better agree with the global X-ray spectrum of Tycho's SNR with respect to deflagration models, but the origin of the supernova explosion is still debated.  
Measuring the mass ratios between intermediate mass elements (IMEs) and iron group elements (IGEs), \cite{2025ApJ...986..216H} compared the observational results obtained with \textit{Suzaku} with different models for the progenitor system and explosion mechanism, setting constraints on the exploding WD mass and on the explosion mechanism. In their case, the IME-to-IGE mass ratios are in agreement with both near- and sub-M$_{Ch}$ progenitor and with the D$^6$ models considering a fraction of 75\% to 95\% of shocked ejecta.
The ejecta emission can indeed provide information on the processes governing the explosion \citep{2008ApJ...680.1149B,2006A&A...453..567M}, especially those associated with the Fe-group (such as Ti, Cr, Mn and Ni) which are synthesized in the inner layers of the exploding star \citep{2008ApJ...680.1149B,2014ApJ...785L..27Y}. 
Chromium and Manganese have been detected in Tycho's SNR with \textit{Suzaku} \citep{2009PASJ...61S.167T,2014ApJ...780..136Y, 2025ApJ...986..216H} and \textit{XMM-Newton} \citep{2015ApJ...805..120M}. Their line centroids were found at lower energies than those characteristic of He-like emission, suggesting an ionization state below the helium-like level \citep{2009PASJ...61S.167T} and similar to that of Fe-K \citep{2009PASJ...61S.167T}.

The high abundance of Mn in Tycho's SNR \citep{2008ApJ...680.1149B}, and also in Kepler's SNR \citep{2013ApJ...767L..10P} suggested a progenitor with super-solar metallicity (population I progenitor). However, even the supernova remnant N103B in the Large Magellanic Cloud (LMC) shows high Mn abundance \citep{2013ApJ...766...44Y}, while a progenitor with solar metallicity in the LMC is improbable \citep{2017ApJ...843...35M}, therefore this suggests that the Mn abundance has been affected by either carbon simmering or electron-capture during the explosion itself \citep{2016sros.confE.105B}.

In this paper we take advantage of the high spectral resolution of \textit{XRISM}/Resolve to determine the spectral properties of chromium and manganese and their spatial variations in Tycho's SNR. The unprecedented spectral resolution of \textit{XRISM}/Resolve is indeed capable of detecting faint emission lines enabling much more precise spatial distribution of Cr and Mn with respect to previous works. We use these new determinations of Cr and Mn in Tycho's SNR to investigate its SN progenitor system and explosion mechanism by comparing to a set of 1D, 2D and 3D Type Ia SN models.

The paper is organized as follows.
In Sect. \ref{sec:reduction} we describe the data reduction, data analysis and results are presented in Sect. \ref{sec:analysis}. The results are discussed in Sect.\ref{sec:discussion}. Finally, conclusions are given in Sect. \ref{sec:conclusions}.

    \begin{figure}[ht!]
    \centering
    \includegraphics[width=\columnwidth]{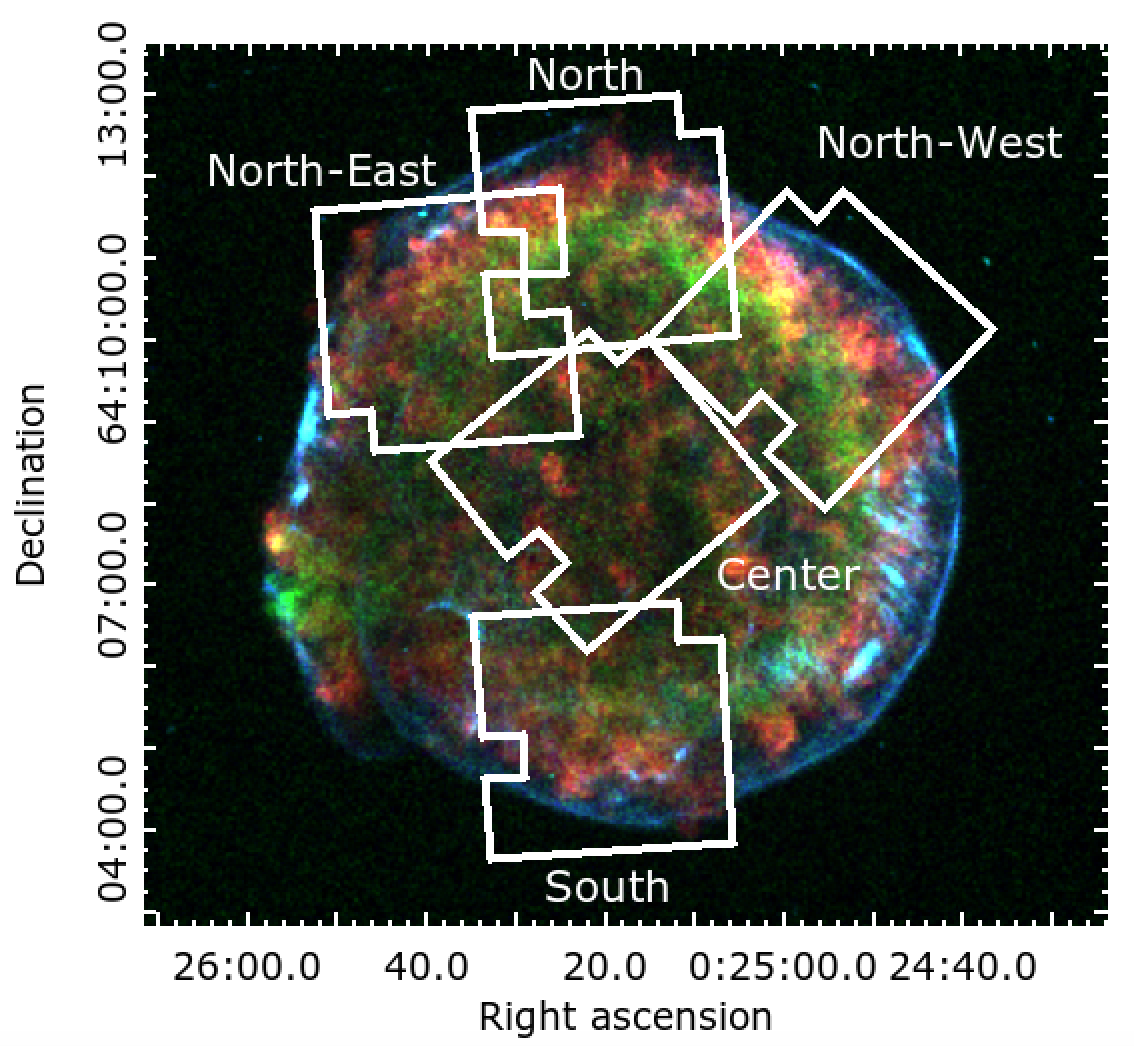}
    \caption{Composite 3-colour image from \textit{Chandra} observations in the 1.5-2.3 keV energy range in red, 4.0-6.0 keV in blue, 6.0-7.0 keV in green. Regions in white show all the analyzed pointings, without pixel \#12 (calibration pixel) and \#27 (unstable pixel, see Sect. \ref{sec:reduction}).}
    \label{fig:obs}
    \end{figure}

\section{Observations and data reduction}
\label{sec:reduction}

\begin{table*}[ht!]
        \caption{Observations parameters}
        \begin{center}
                \footnotesize
                \begin{tabular}{c c c c c c}
                    \hline\hline
                        Name & Obs ID & RA (deg)  & DEC (deg) & Roll Angle (deg) & Texp (ks) \\
                        \hline
                        NW &000160000 & 6.23 & 64.17 & 226.37 & 204 \\
                        C &300060010 & 6.33 & 64.14 & 219.36 & 175 \\
                        N & 300062010 & 6.33 & 64.19 & 184.87 & 42 \\
                        S & 300063010 & 6.33 & 64.09 & 183.69 & 46 \\
                        NE & 201118010 & 6.41 & 64.17 & 4.89 & 100 \\
            \hline
                \end{tabular}
                \label{tab:obs}
        \end{center}
\end{table*}

We analyzed observations on different regions of Tycho's SNR performed with \textit{XRISM}/Resolve during the Performance Verification (PV) phase, namely North-West (NW, ObsID 000160000), Center (C, ObsID 300060010), North (N, ObsID 300062010), South (S, ObsID 300063010) and one observation performed during AO1 phase, namely North-East (NE, ObsID 201118010). Figure \ref{fig:obs} shows the five analyzed pointings overlaid on a \textit{Chandra} 3-colour composite map. All useful information regarding the observations is shown in Tab. \ref{tab:obs}. 

For data reduction and spectral analysis we followed the guide provided by the \textit{XRISM} consortium\footnote{\url{https://heasarc.gsfc.nasa.gov/docs/xrism/analysis/abc_guide/xrism_abc.pdf}}.
All observations were reprocessed with
the \textit{HEASoft} 6.34 software package, applying calibration from \textit{XRISM} \texttt{CalDB} 9 (Version 20240815). We applied the default screening as described in \cite{2024PASJ...76.1186X}.  

Only high-resolution primary (‘Hp’ or ITYPE=0) grade events were included. We always exclude pixel \#27, which has been shown to exhibit erratic gain jumps. 
The redistribution matrix files (RMF) were generated in the large (L) size using the \texttt{rslmkrmf} task. The anciliary response files (ARF) were built with \texttt{xaarfgen} task, using two SNR images generated with general morphological component analysis (GMCA, see Appendix \ref{sec:app_GMCA}), as an input model for source flux (as described in Sect. \ref{sec:analysis}). The Non-X-ray Background (NXB) spectrum was modeled from a database\footnote{\url{ https://heasarc.gsfc.nasa.gov/docs/xrism/analysis/nxb/index.html}} of Resolve using \texttt{rslnxbgen}.
The NXB has been always determined from the global pointing (the entire array except pixel \#12 and \#27) for statistical reasons and, when moving to smaller regions, it has been rescaled for the surface ratio.

\section{Data analysis and results}
\label{sec:analysis}

We used \texttt{Xspec} version 12.14.1, with AtomDB version 3.1.3. We set the parameter \texttt{APECBROADPSEUDO} in the 'On mode' in order to allow the broadening of the “pseudo-continuum” lines. Due to the Poissonian distribution of the data, we fit the spectra using the C-statistic \citep{Cash79}.
As the objective is to study the Cr and Mn emission, we focus our analysis to the energy range 4-15 keV, dominated by the Fe group elements and nonthermal continuum emission. 
The lower limit of the energy range has been selected to exclude contribution from significant emission lines at lower energies (such as Calcium, Argon etc.). The upper limit is the maximum reliable value for the current NXB model provided by the consortium (see Sect. \ref{sec:reduction}). It is high enough to have a good handle on the non-thermal continuum spectral shape. In this energy range, there are two different spectral components with different morphologies: the non-thermal emission (synchrotron emission) and the thermal emission coming from the Fe-group shocked ejecta. \cite{2015ApJ...805..120M} indeed have shown that Cr and Mn are spatially correlated with Fe. In order to take into account these different spatial/spectral components, we use a multiple-response methodology, associating different ARFs to different (spatial and spectral) components. The ARF generated with the synchrotron spatial distribution as input image (right panel in Fig. \ref{fig:gmca}) is associated to the non-thermal component of the model. On the other hand, the ARF generated with the Fe spatial distribution (left panel in Fig. \ref{fig:gmca}) as input image is associated to the Fe-group ejecta thermal component. Using different ARFs for the iron group ejecta and for the nonthermal emission helps accounting for the different contamination of the emission from the neighbouring regions, due to the point spread function (PSF) of about 1.7 arcmin (effect of the spatial spectral mixing, whose study is beyond the scope of this paper). 

    \begin{figure*}[ht!]
    \centering
    \includegraphics[width=\textwidth]{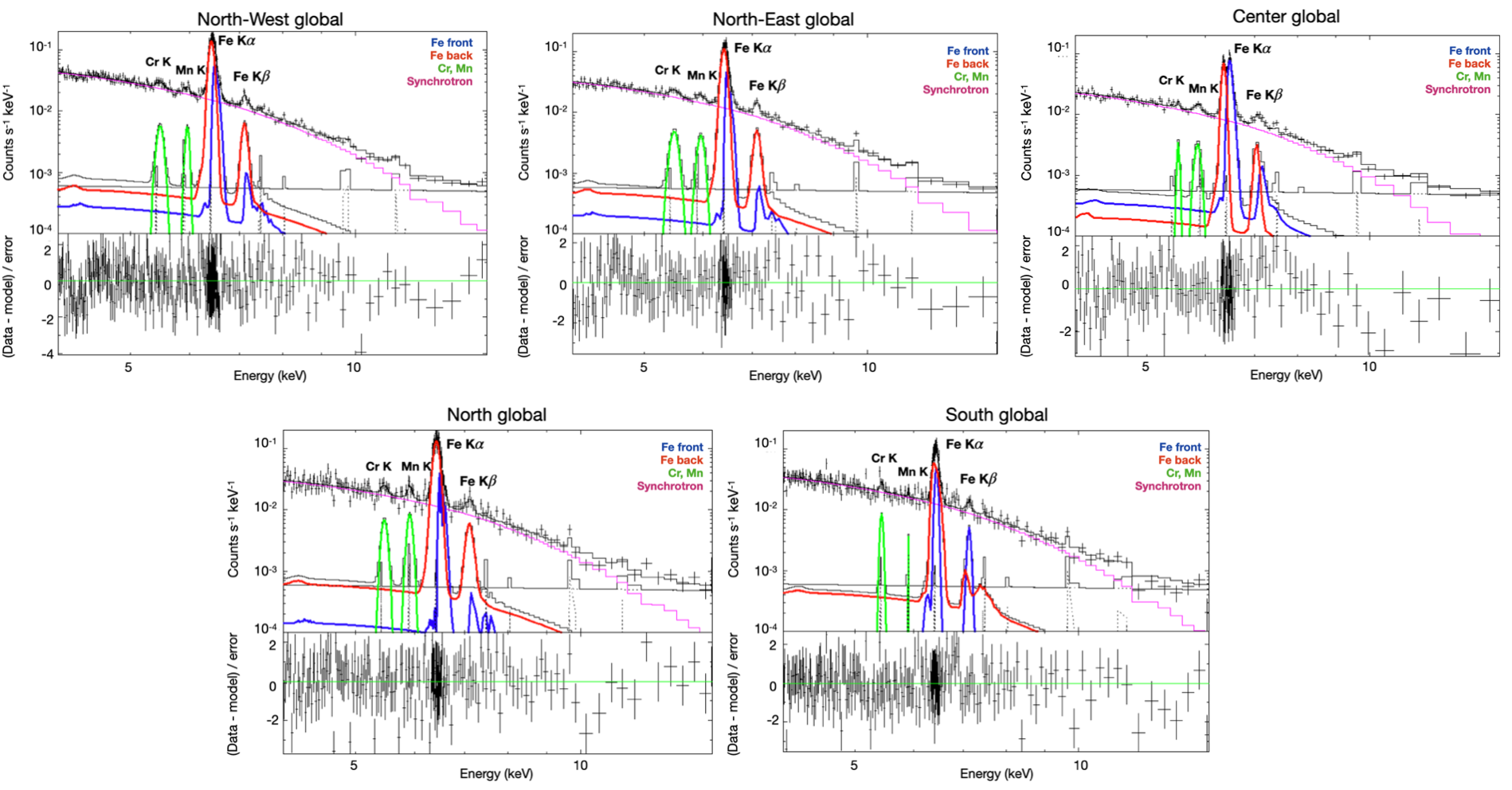} 
    \caption{XRISM/Resolve global spectra for each pointing (shown in Fig. \ref{fig:obs}), with their associated best fit model (defined in Eq. \ref{eq:mod_gauss}). Blue and red components show the blueshifted and resdshifted Fe emissions, green components show the Gaussian lines for Cr-K and Mn-K emission lines respectively. The powerlaw component is shown in magenta.}
    \label{fig:Es_spec_gauss}
    \end{figure*}

\subsection{Detection and spatial distribution of Cr and Mn emission lines}
\label{sec:gauss}
In the first part of the analysis, we use Gaussian line models to determine where Cr and Mn emission lines are significantly detected both in the overall spectrum of each observed pointing, and on a smaller spatial scale. We adopt the following model, composed of both thermal and nonthermal components, to fit the spectrum between 4 and 15 keV and quantify the level of Cr and Mn line emission.

\begin{equation}
\label{eq:mod_gauss}
\begin{split}
    Model& = \frac{1}{c_1}[bvvpshock_{rear} \; (\text{Fe})  + bvvpshock_{front} \; (\text{Fe}) + \\ & + Gauss(Cr) + Gauss(Mn)]\footnotemark \\
    & + \frac{1}{c_2}[Power Law]\footnotemark
\end{split}
\end{equation}
\addtocounter{footnote}{-1} 
\footnotetext{The associated ARF is generated with the Fe image (upper panel of Fig. \ref{fig:gmca})}
\stepcounter{footnote}
\footnotetext{The associated ARF is generated with the synchrotron image (lower panel of Fig. \ref{fig:gmca})}

where $bvvpshock$ (\texttt{Xspec} package) is the constant temperature plane-parallel non equilibrium ionization shock plasma model \citep{1983ApJS...51..115H,1994ApJ...429..710B,1995ApJ...438L.115L,2001ApJ...548..820B}. The abundances in the $bvvpshock$ model are relative to their solar values taken from \cite{2000ApJ...542..914W}. Two separated blueshifted and redshifted thermal components (modelled by $bvvpshock$), are required to reproduce the front and rear Fe shocked ejecta shell, as revealed by kinematic studies with \textit{XRISM}/Resolve (XRISM collaboration 2026, in prep and \cite{2026A&Asubn}.). The two $Gauss$ are the Gaussian line profiles centered at the Cr-K and Mn-K emission lines. The power law component accounts for the nonthermal synchrotron emission from TeV electrons accelerated at the shock, which is the dominant contribution to the high energy continuum. The two constants, $c_1$ and $c_2$, are the flux ratios between the considered region and the whole SNR in the input images (see Appendix \ref{sec:app_GMCA}) used for the ARF calculation. After taking into account those two constants, the normalizations of the components correspond to the fluxes in the selected regions.

For the thermal $bvvpshock$ components, which reproduce the Fe-K emission, we made the following assumptions. As the electron temperature of the iron plasma is not well constrained because of the dominant synchrotron continuum with respect to the thermal bremsstrahlung, we fixed it at 10 keV (as in \citealt{2025A&A...693A.234G}). Similar issues has been found in the \textit{XRISM}/Resolve spectral analysis of Cas A \citep{2026arXiv260206952A}. Our Cr and Mn flux results appear not affected by the temperature of free electrons in the Fe plasma, as already found by \cite{2014ApJ...780..136Y}. To account for a pure ejecta heavy element composition, the Fe abundance was set at 1000, as in \cite{2020A&A...638A.101G}. All other abundances are fixed at 0. Ionization time, redshift, velocity dispersion and normalization, of the front and rear Fe plasma components, are free to vary.
The two Gaussian lines account, when required, for the emission from Cr and Mn. 
We claim the detection of Cr and Mn only where the line flux is significant to at least 3$\sigma$ level. This corresponds to $\Delta$C-stat higher than 14 for the modelling of the global spectra (three free parameters, Tab. \ref{tab:global_gauss}) and $\Delta$C-stat higher than 12 for the spectral modelling of subregions (two free parameters, Tab. \ref{tab:gauss}). Less stringent constraints are provided by the 90\% confidence error bars. 

The spectral analysis has first been conducted at the size of the global pointings. Fig. \ref{fig:Es_spec_gauss} shows the global spectra extracted from each pointing, in the 4.0-15.0 keV band with their best fit model.
Best fit parameters associated with the Gaussians and the power law are shown in Tab. \ref{tab:global_gauss} (error bars are provided at 90\% confidence level). 
The analysis shows significant emission at 3 sigma of Cr only in the NW and NE, while Mn is in all pointings except the south. Fluxes of both Cr and Mn in the NW pointing are in good agreement with previous results obtained with \textit{Suzaku} in the same area \citep{2014ApJ...780..136Y}.
As a result, Cr and Mn are detected where iron emission is the brightest, in the NE and NW rim, and only Mn and at a fainter level in the center and in the north. Note that at 90\% confidence level, Cr is also detected in the north, south and center. However, the low statistics in the north and in the south, due to limited observation time, prevent any firm conclusion.

\begin{figure}[h!]
    \centering
    \includegraphics[width=\columnwidth]{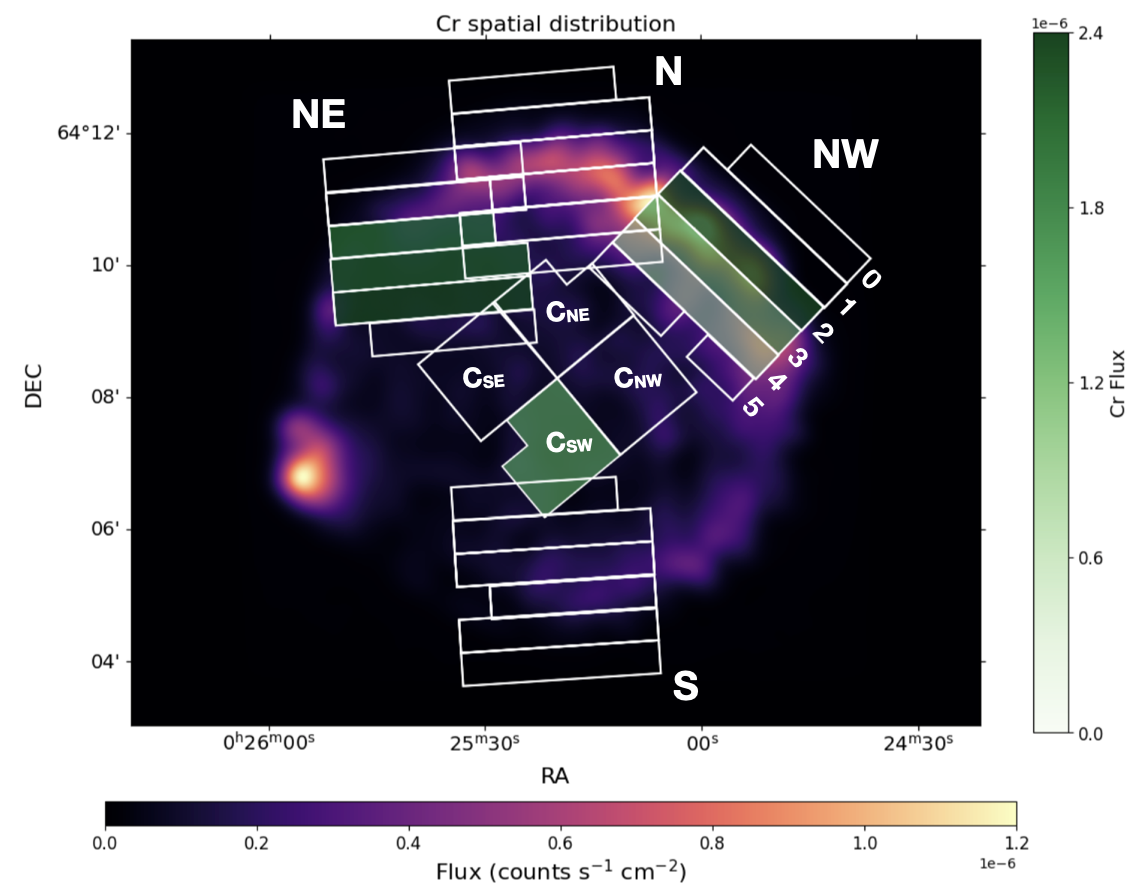}
    \includegraphics[width=\columnwidth]{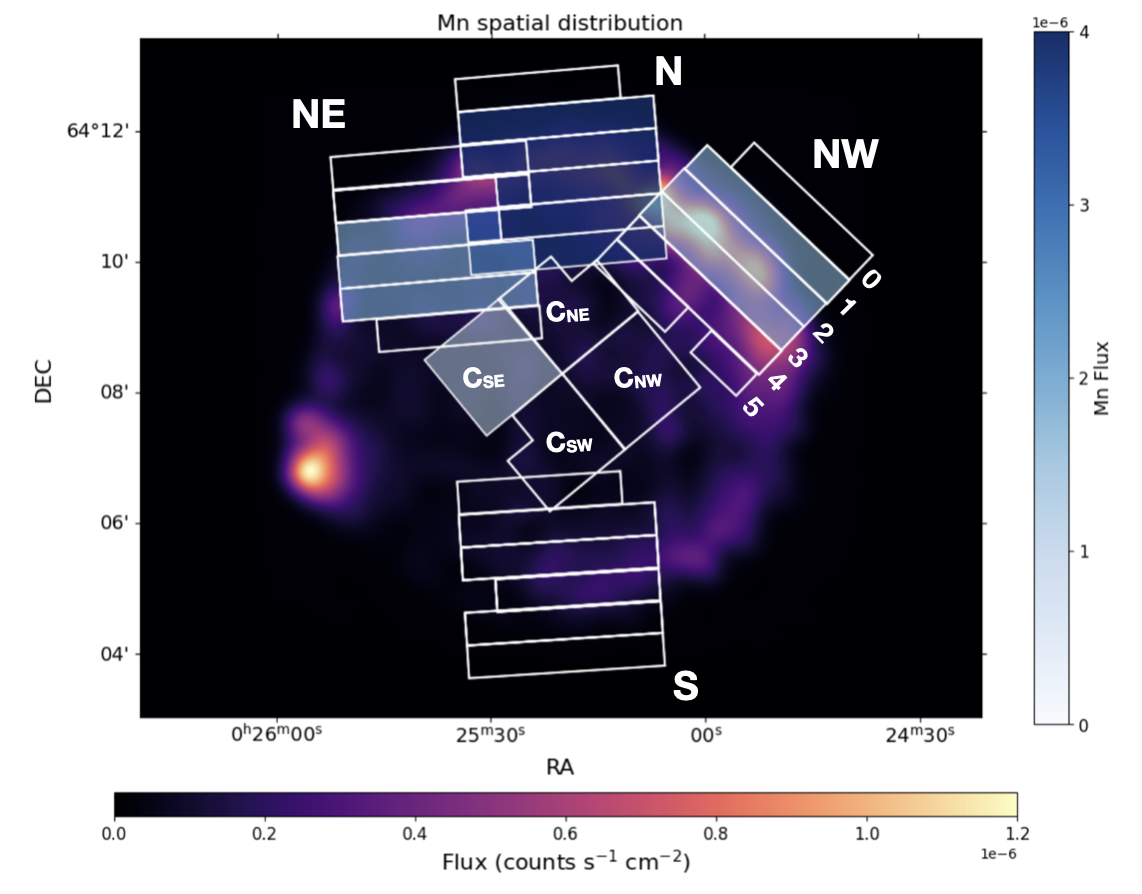}
    \caption{Cr and Mn fluxes are shown in the shaded regions in green in the upper panel and blue in the lower panel, respectively. They are overlaid on the \textit{Chandra} Fe image obtained with GMCA (see Appendix \ref{sec:app_GMCA}) convolved with a Gaussian with the same \textit{XRISM}/Resolve PSF, showing the Fe emission. We caution the reader that there are some overlaps between rows (e.g. row 2 is included in both NW$_{12}$ and NW$_{23}$). In the intermediate rows we show the mean flux between the two overlapping regions.}
    \label{fig:spatial}
    \end{figure}

In order to provide a more precise spatial distribution of Cr and Mn, we analyzed spectra from smaller regions (see Fig. \ref{fig:spatial}), following a radial profile for the shell pointings (NW, NE, N, S), and groups of pixels for the center. Indeed, most of the explosion mechanisms predict a radial stratification of the elements (see Sect. \ref{sec:discussion}). For the NW and NE pointings we extracted the spectra from two rows of pixels (12 pixels in total) to have enough statistics for the detection of such faint emission lines. For the North and the South, as the exposure time and then the statistics is lower, we need at least 3 rows (18 pixels in total). For the center, groups of at least 9 pixels are required. The broadening (Sigma) of the Gaussian lines is fixed to the best-fit value obtained from the modelling of the global spectrum of each pointing (see Tab. \ref{tab:global_gauss}). Regions with significant emission for Cr and Mn are shown, overlaid on the iron map generated with GMCA (see Appendix \ref{sec:app_GMCA}) convolved with a Gaussian with \textit{XRISM}/Resolve PSF, in Fig. \ref{fig:spatial}. The best fit parameters of each sub-region are shown in Tab. \ref{tab:gauss}. Spectra and their associated best fit models are shown in Appendix \ref{sec:app_spectra}.

    \begin{table*}[h!]
\centering
        \caption{Best fit parameters of the Gaussian associated with Cr and Mn and the power law when fitting with Eq. \ref{eq:mod_gauss}. Errors are at 90\% confidence level.}
                \footnotesize
                \begin{tabular}{c c c c c c c c c c c c c}  
                    \hline\hline
                    & & Cr & & & & Mn & & & & & Pow\\
                    \hline
                    Reg & c$_1$ & Centroid  & Sigma& N ($\times 10^{-6}$) & $\Delta C^{(a)}$ & Centroid & Sigma & N ($\times 10^{-6}$) & $\Delta C^{(a)}$  & c$_2$ & Index & N$^{(c)}$ ($\times10^{-2}$)\\
                    & & (keV) & (keV) & (ph s$^{-1}$cm$^{-2}$) & & (keV) & (keV) & (ph s$^{-1}$cm$^{-2}$) & & & & (ph keV$^{-1}$cm$^{-2}$)\\
                    \hline
                    \vspace{0.1cm}
                    NW & 0.25 & $5.48^{+0.03}_{-0.03}$ & $0.05^{+0.03}_{-0.02}$ & $4^{+2}_{-2}$ & 25 & $5.96^{+0.03}_{-0.02}$ & $0.05^{+0.03}_{-0.02}$ & $2.7^{+1.1}_{-1.2}$ & 19 & 0.18 & $2.96^{+0.04}_{-0.04}$ & $2.1^{+0.2}_{-0.2}$\\
                    \vspace{0.1cm}
                    N & 0.21 & $5.46^{+0.04}_{-0.04}$ & $0.05^{+0.03}_{-0.02}$ & $5^{+3}_{-3}$ & 9 & $5.91^{+0.03}_{-0.03}$ & $0.05^{+0.03}_{-0.02}$ & $6^{+3}_{-3}$ & 17 & 0.14 &$2.76^{+0.12}_{-0.12}$ & $1.1^{+0.3}_{-0.2}$\\
                    \vspace{0.1cm}
                    NE & 0.25 & $5.49^{+0.04}_{-0.04}$ & $0.06^{+0.04}_{-0.02}$ & $5^{+2}_{-2}$& 21 & $5.96^{+0.04}_{-0.04}$ & $0.06^{+0.05}_{-0.03}$ & $4^{+2}_{-2}$ & 17 & 0.18 &$2.85^{+0.07}_{-0.07}$ & $1.4^{+0.2}_{-0.2}$\\
                    \vspace{0.1cm}
                    C & 0.09 & $<5.54$ & $0.02^{+0.16}_{-0.01}$ & $1.1^{+2.8}_{-0.9}$ & 7 & $5.86^{+0.03}_{-0.04}$ & $0.05^{+0.03}_{-0.02}$ & $2.6^{+1.5}_{-1.3}$ & 15 & 0.08 &$2.96^{+0.08}_{-0.08}$ & $1.1^{+0.1}_{-0.2}$\\
                    \vspace{0.1cm}
                    S & 0.07 & $5.43^{+0.04}_{-0.04}$ & $0.02^{+0.03}_{-0.01}$ & $3^{+2}_{-2}$ & 7 &  N.C.$^{(b)}$  &  N.C.$^{(b)}$ &  N.C.$^{(b)}$ & 1 & 0.11 & $2.91^{+0.11}_{-0.11}$ & $1.8^{+0.3}_{-0.3}$\\
                        \hline
            \hline
                \end{tabular}
                $^{(a)}$ $\Delta$C$_{\rm{stat}}$ calculated as the difference of the  C$_{\rm{stat}}$ of the model without and with the line. \\ $^{(b)}$ Not Constrained (N.C.) or unreasonable values for lower and upper limits. \\ $^{(c)}$ At 1 keV
                \label{tab:global_gauss}
\end{table*}

We also checked for the presence of Ni-K$\alpha$ emission line (centered at $\sim$ 7.5 keV) in the spectra, observed in the \textit{Suzaku} data analysis \citep{2014ApJ...785L..27Y,2025ApJ...986..216H}. We do not detect any significant emission lines at Ni-K$\alpha$ energy range at 3$\sigma$ confidence level, in any of the observed regions. The high spectral resolution of \textit{XRISM}/Resolve can indeed better disentangle Ni-K$\alpha$ emission from background lines and very close emission lines, such as Fe-K$\beta$.

In conclusion, we show that most of the emission of Cr and Mn is located along the bright Fe rim and that the emission significantly varies at smaller scale along the NE-NW Fe rim. Our spatially resolved spectral analysis provides the spatial distribution of Cr and Mn along the NW and NE rim, and in the center where it is more difficult to detect. Despite the lower statistics, the Mn line is also detected and spatially resolved in the north, but only with large regions of 3 rows. The S observations have too low exposure to allow a 3 sigma detection of these faint lines, however the Mn line is significant at 90\% confidence level. 

\subsection{Physical modelling for Cr, Mn and mass ratios}
\label{sec:mass}

\begin{figure}[h]
    \centering
    \includegraphics[width=\columnwidth]{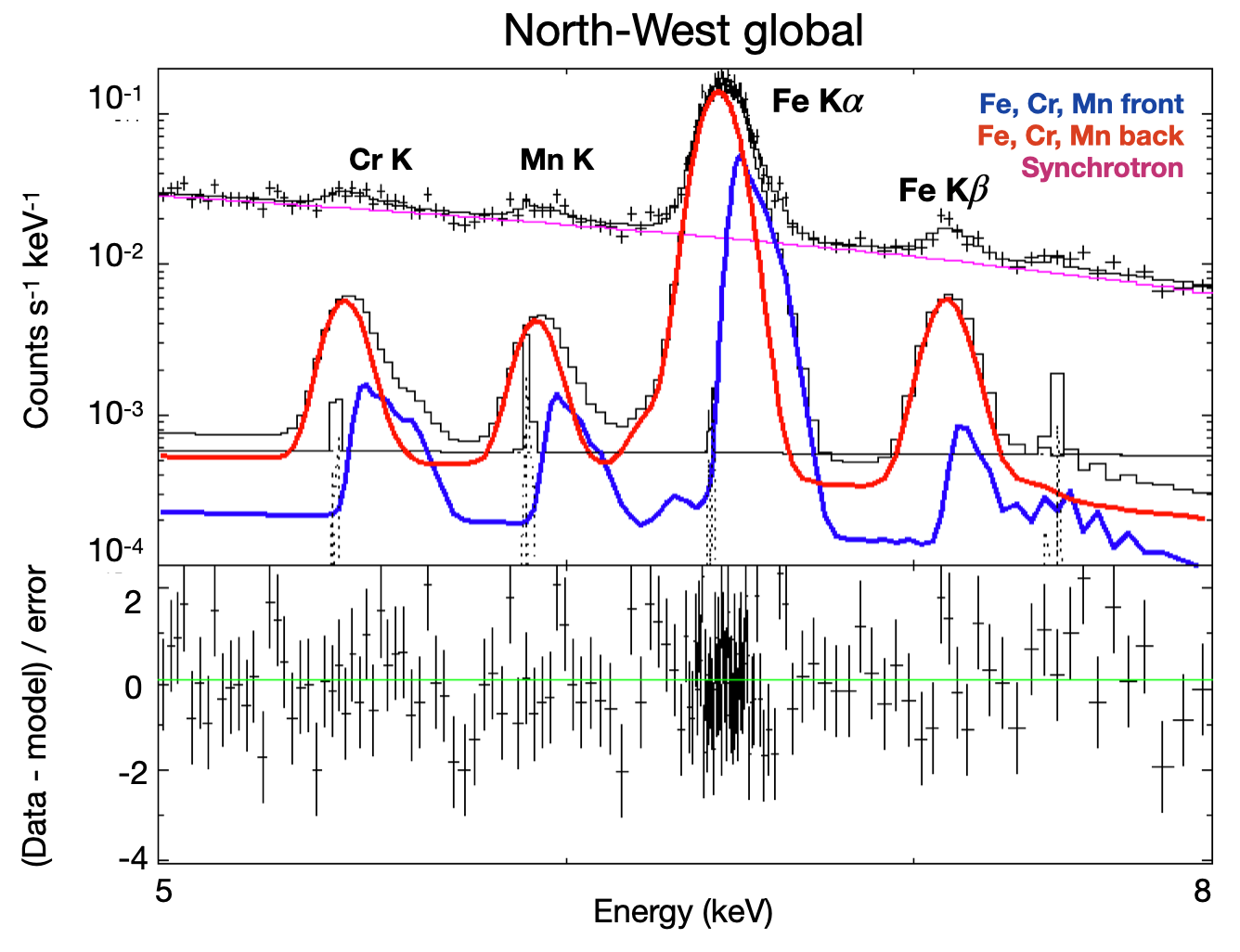}
    \caption{XRISM/Resolve best fit global spectrum of the NW pointing with the model in Eq. \ref{eq:mod_pshock}. Blue and red lines show the blueshifted and resdshifted Fe-group plasma components, respectively. The powerlaw continuum is shown in magenta. The NXB, which consists of a powerlaw plus Gaussians (dotted lines), is shown in black and its flux is lower than $10^{-3}$ counts s$^{-1}$ keV$^{-1}$.}
    \label{fig:Es_spec_pshock}
    \end{figure}

In the second part of the analysis, we perform a physical modelling of the global spectra per pointing, where both Cr and Mn were significantly detected, and determine the properties of the Cr and Mn emitting plasma. In addition to be spatially correlated (Sect. \ref{sec:gauss}), Cr and Mn have been found to share the same physical properties as Fe \citep{2015ApJ...805..120M,2025ApJ...986..216H}. We then consider that the plasma emitting Cr and Mn is associated with the Fe plasma and shares the same properties as those of the front and rear Fe components. We thus adopt the following model:

\begin{equation}
\label{eq:mod_pshock}
\begin{split}
    Model &= \frac{1}{c_1}[bvvpshock_{rear} \; (\text{Cr,Mn,Fe}) + \\ & + bvvpshock_{front} \; (\text{Cr,Mn,Fe})]^3 + \frac{1}{c_2}[Power Law]^4
\end{split}
\end{equation}

where the Fe abundance is frozen to 1000 in both the front and rear components (as for the previous model in Sect. \ref{sec:gauss}) and the abundances for Cr and Mn are free to vary with the same front/rear abundance for each element.
Tab. \ref{tab:mass} provides the resulting best-fit parameters for the physical model in Eq. \ref{eq:mod_pshock} of the global spectra for each pointing. For visualization purposes, Fig. \ref{fig:Es_spec_pshock} shows a close-up view of the NW spectrum, in the 5.0-8.0 keV with the best-fit model. All other spectra are shown in Appendix \ref{sec:app_spectra}.
In all cases, we need a range of ionization times ($\tau$) for both blueshifted and redshifted Fe components, with the lower limit ($\tau_l$) at 0. 

Due to the dominant synchrotron emission with respect to the thermal ejecta continuum, our X-ray spectral fits show a degeneracy between emission measure (associated with the normalization in the \texttt{xspec} $bvvphsock$ models) and absolute abundances. Because of the unknown filling factor along the line of sight, the determination of the absolute mass of individual elements is not always reliable. However we can have a good handle on the mass (or abundance) ratios \citep{2020A&A...638A.101G}. 
We then measured the relative mass of Cr and Mn with respect to the Fe. 
We calculate the mass ratios through the following equation:

\begin{equation}
\centering
    \frac{M_X}{M_{Fe}} = \frac{Ab_X}{Ab_{Fe}} \frac{[X/H]_\odot} {[Fe/H]_\odot} \frac{m_{atom_X}}{m_{atom_{Fe}}} 
\end{equation}

where $Ab_X$ is the abundance of element X relative to hydrogen over its solar ratio, ${[X/H]_\odot}$ is the corresponding solar abundance \citep{2000ApJ...542..914W}, $m_{atom_X}$ is the atomic mass number of element X. The resulting mass ratios of Cr and Mn over Fe, for each pointing, are given in Tab. \ref{tab:mass}. Error bars are at 90\% confidence level. 
In order to increase the statistics, we also measured the average mass ratios through a simultaneous spectral analysis of all the four pointings (NW, NE, C, N). In this case all the physical parameters of each spectrum are free to vary, but the overall abundances of Cr and Mn are forced to be the same for all spectra. For this global analysis, we obtain $M_{\rm{Cr}}/M_{\rm{Fe}} = 0.026 \pm 0.006$, $M_{\rm{Mn}}/M_{\rm{Fe}} = 0.026 \pm 0.007$. The spectra and the best-fit model are shown in Fig. \ref{fig:global_spec_average}.
With this analysis we also derived the Mn/Cr mass ratio, which allows us to measure the neutron excess of the progenitor system (see Sect. \ref{sec:disc_met}), obtaining $M_{\rm{Mn}}/M_{\rm{Cr}} = 1.0 \pm 0.3$.

While previous works observed Ni in their spectral analysis \citep{2025ApJ...986..216H,2014ApJ...780..136Y}, Ni is not found significant in any of the analyzed \textit{XRISM}/Resolve spectra. However, the upper limit at 3$\sigma$ confidence level of the Ni/Fe mass ratio, for the global analysis, is found to be $M_{\rm{Ni}}/M_{\rm{Fe}} = 0.04$, consistent with the upper value of the \textit{Suzaku} measurement ($M_{\rm{Ni}}/M_{\rm{Fe}} = 0.025 \pm 0.015$) from \cite{2025ApJ...986..216H}.

\begin{table*}[h!]
\centering
        \caption{Best fit parameters obtained by fitting the regions with the model in Eq. \ref{eq:mod_pshock}. Errors are at 90\% confidence level.}
                \footnotesize
                \begin{tabular}{cccccccc}
                    \hline\hline
                    & & & & \texttt{Fe, Cr, Mn front} & & & \\
                    \hline
                    Reg & c$_1$ & Ab$_{Cr}$ & Ab$_{Mn}$ & $\tau_u$  & Doppler shift & Vel Disp & N$^{a}$  \\
                    & & ($\times 10^3$) & ($\times 10^3$) & ($\times 10^{10}$) s cm$^{-3}$ & ($\times10^{-3}$) & ($\times10^{3}$) km s$^{-1}$ & ($\times10^{-5}$) cm$^{-5}$ \\
                    \hline
                    \vspace{0.1cm}
                    NW & 0.25 & $2.4_{-0.9}^{+0.9}$ & $2.9_{-1.3}^{+1.3}$ & $1.6^{+0.3}_{-0.2}$ & $-7.7^{+0.8}_{-1.8}$ & $0.9^{+0.4}_{-0.4}$ & $0.9^{+0.4}_{-0.4}$\\
                    \vspace{0.1cm}
                    N & 0.21 & $2.6^{+1.7}_{-1.6}$& $5.6_{-2.6}^{+2.8}$ & $2.1^{+0.9}_{-0.6}$ & $-6.9^{+0.5}_{-0.5}$ & $<0.5$ & $0.5^{+0.7}_{-0.2}$ \\
                    \vspace{0.1cm} 
                    NE & 0.25 & $2.5_{-1.1}^{+1.2}$ & $3.3_{-1.6}^{+1.7}$ &$1.9^{+0.5}_{-0.2}$ & $-4.2^{+0.6}_{-0.6}$ & $0.4^{+0.2}_{-0.2}$ & $0.7^{+0.4}_{-0.4}$\\
                    \vspace{0.1cm}
                    C & 0.09 &$2.0_{-1.6}^{+1.7}$ & $2.7_{-1.7}^{+1.8}$ & $1.2^{+0.2}_{-0.2}$ & $-6.4^{+0.9}_{-0.9}$ & $1.4^{+0.4}_{-0.3}$ & $1.5^{+0.1}_{-0.1}$\\
            \hline
            \hline
            & & & & \texttt{Fe, Cr, Mn rear} & & & \\
             \hline
             Reg & & Ab$_{Cr}$ & Ab$_{Mn}$ & $\tau_u$  & Doppler shift & Vel Disp & N$^{a}$ \\
             & & & & ($\times10^{10}$) s cm$^{-3}$ & ($\times10^{-3}$)& ($\times10^{3}$ km $s^{-1}$& ($\times10^{-5}$) cm$^{-5}$ \\
             \hline
             NW & & Linked to front& Linked to front& $0.5^{+0.3}_{-0.1}$ & 0 & $2.1^{+0.1}_{-0.1}$ & $3.2^{+0.3}_{-0.3}$ \\
              \vspace{0.1cm}
              N & &Linked to front&Linked to front& $0.5^{+0.6}_{-0.2}$ & $<4.4$ & $2.7^{+0.2}_{-0.6}$ & $3.8^{+0.2}_{-0.8}$ \\
              \vspace{0.1cm}
             NE & & Linked to front& Linked to front & $0.5^{+0.3}_{-0.2}$ & $1.8^{+1.8}_{-1.5}$ & $2.4^{+0.2}_{-0.3}$ & $2.9^{+0.4}_{-0.4}$\\
              \vspace{0.1cm}
             C & & Linked to front& Linked to front& $0.4^{+0.3}_{-0.1}$ & $10^{+2}_{-1}$ & $1.7^{+0.2}_{-0.2}$ & $1.4^{+0.2}_{-0.2}$\\
             \hline
             \hline
             & & & Power Law \\
             Region &  $M_{\rm{Cr}}/M_{\rm{Fe}}$ & $M_{\rm{Mn}}/M_{\rm{Fe}}$ &c$_2$ & Photon Index  & N$^{a}$ \\
              & && & & (ph keV$^{-1}$ cm$^{-2}$) \\
             \hline
             NW & $0.03^{+0.010}_{-0.010}$ & $0.02^{+0.01}_{-0.01}$ & 0.18 & $2.97^{+0.05}_{-0.05}$ & $2.1^{0.2}_{0.2}\times10^{-2}$ & \\
             \vspace{0.1cm}
             N & $0.03^{+0.02}_{-0.02}$ & $0.04^{+0.02}_{-0.02}$ &  0.14 &$2.8^{+0.1}_{-0.1}$ & $1.1^{+0.3}_{-0.2}\times10^{-2}$\\
              \vspace{0.1cm}
             NE & $0.03^{+0.01}_{-0.01}$ & $0.03^{+0.01}_{-0.01}$ &  0.18 &$2.87^{+0.07}_{-0.07}$ & $1.4^{+0.2}_{-0.2}\times10^{-2}$ \\
              \vspace{0.1cm}
             C & $0.023^{+0.018}_{-0.020}$ & $0.02^{+0.01}_{-0.01}$ &  0.08 & $2.98^{+0.08}_{-0.08}$ & $1.1^{+0.1}_{-0.2}\times10^{-2}$\\
            \hline
            \hline
                \end{tabular}
                \label{tab:mass}
                \\ $^{(a)}$ At 1 keV
\end{table*}

\section{Discussion}
\label{sec:discussion}
The spatial distribution and the mass of Cr and Mn carry important information to study the nucleosynthesis and explosion mechanism of SNe.
These faint emission lines have been previously detected in the spectral analysis of Tycho's SNR globally with \textit{Suzaku} \citep{2009PASJ...61S.167T,2025ApJ...986..216H} and in selected regions with Fe emission with \textit{XMM-Newton} \citep{2015ApJ...805..120M} and also \textit{Suzaku} \citep{2014ApJ...780..136Y,2017ApJ...834..124Y}. 
The unprecedented spectral resolution of \textit{XRISM}/Resolve allowed us to further constrain the Cr and Mn emission and measure their spatial distribution at smaller scales. As shown in Sect. \ref{sec:analysis} and Fig. \ref{fig:spatial}, we find significant emission of both Cr and Mn in the North-East and North-West, where there is bright Fe emission and sufficient exposure time of the observations (Tab. \ref{tab:obs}). The observation in the Center has also very good exposure time, but the iron emission is much fainter there and the global spectrum shows only Mn emission and no significant Cr emission line at 3 $\sigma$ (significant at 90\% confidence level). The much smaller exposure of the observations in the North and in the South (around 40 ks each) significantly affects the detection of these faint emission lines. We find significant Mn emission in the northern rim of the remnant, but no Cr is detected at 3 $\sigma$ level (it is at 90\% confidence level). In the southern rim, neither Cr nor Mn are detected at 3 $\sigma$ level (but Cr is at 90\% confidence level).

The global Cr/Fe and Mn/Fe mass ratios in Tycho's SNR have been previously determined in \cite{2008ApJ...680L..33B,2017ApJ...843...35M,2025ApJ...986..216H}. With \textit{XRISM}/Resolve, we can estimate the mass ratios at the scale of individual pointings. 
We notice that the mass ratios are well constrained in the NW and NE regions and less well-constrained in the N. For the northern pointing a not very good handle on the mass ratio results from the lack of observation time (see Tab. \ref{tab:obs}). In the central region we integrate over a longer line of sight and expect two different ejecta components, blueshifted and redshifted. The emission lines are therefore expected to be fainter than along the rim. Where Cr and/or Mn are significantly detected (Tab \ref{tab:mass}, Sect. \ref{sec:analysis}), we found that all the mass ratios in different pointings are compatible. While the spatial distribution of the emission line fluxes is not homogeneous, the relative mass ratios of Cr and Mn to Fe appear relatively homogeneous within our error bars.

Our results on Cr/Fe are in good agreement with the global measurements derived from Suzaku observations by \cite{2025ApJ...986..216H} ($M_{\rm{Cr}}/M_{\rm{Fe}} = 0.025 \pm 0.004$) and \cite{2017ApJ...843...35M} ($M_{\rm{Cr}}/M_{\rm{Fe}} = 0.016^{+0.018}_{-0.005}$). On the other hand, for Mn/Fe we find that all mass ratios are significantly higher by about a factor of 2 (see Tab. \ref{tab:mass}) than previously found in \cite{2025ApJ...986..216H} ($M_{\rm{Mn}}/M_{\rm{Fe}} = 0.010 \pm 0.004$) and \cite{2014ApJ...780..136Y} ($M_{\rm{Mn}}/M_{\rm{Fe}} = 0.012^{+0.009}_{-0.004}$).
The high spectral resolution of \textit{XRISM}/Resolve, combined with a well determined background line of Mn, provides a  higher determination of the faint Mn-K emission line which might have been underestimated with previous \textit{Suzaku} data. Using the new version of AtomDB improves the precision of the measurements of Cr and Mn abundances but does not globally change the results.
By modeling individually the spectrum of different regions, we ensure a better handling of the local underlying continuum and line properties. We measured the average mass ratios through a simultaneous spectral analysis of all the four observations, obtaining $M_{\rm{Cr}}/M_{\rm{Fe}} = 0.026 \pm 0.006$, $M_{\rm{Mn}}/M_{\rm{Fe}} = 0.026 \pm 0.007$.
At this stage, we cannot provide the total mass ratio of Cr and Mn to Fe in the shocked ejecta of Tycho's SNR because current \textit{XRISM}/Resolve observations do not cover the entire remnant. 

\subsection{Comparison with different supernova models}
\label{sec:disc_Mch}

\cite{2025ApJ...986..216H} compared their derived global mass ratio between IME and IGEs with a series of various supernova models. They found a good agreement with both near- and sub-M$_{Ch}$ progenitor from \cite{2019MNRAS.482.4346B} and also with 1D D$^6$ models from \cite{2018ApJ...854...52S}. However, their mass ratios are inconsistent with dynamically stable double detonation models in \cite{2018ApJ...854...52S}. 
Considering the mass ratios between IMEs and IGEs, they estimated  $75\%$ to $95\%$ of ejecta have been shocked. 
Their results are in agreement with \cite{2015ApJ...808...49K} who estimated almost 86\% of the ejecta were shocked in Tycho’s SNR by comparing the observed ejecta masses with the total mass predicted by Type Ia supernova models, under the assumption that most of the interior ejecta consist of iron-group elements \citep{2006ApJ...645.1373B}.

\begin{figure*}[h!]
    \centering
    \includegraphics[scale=0.5]{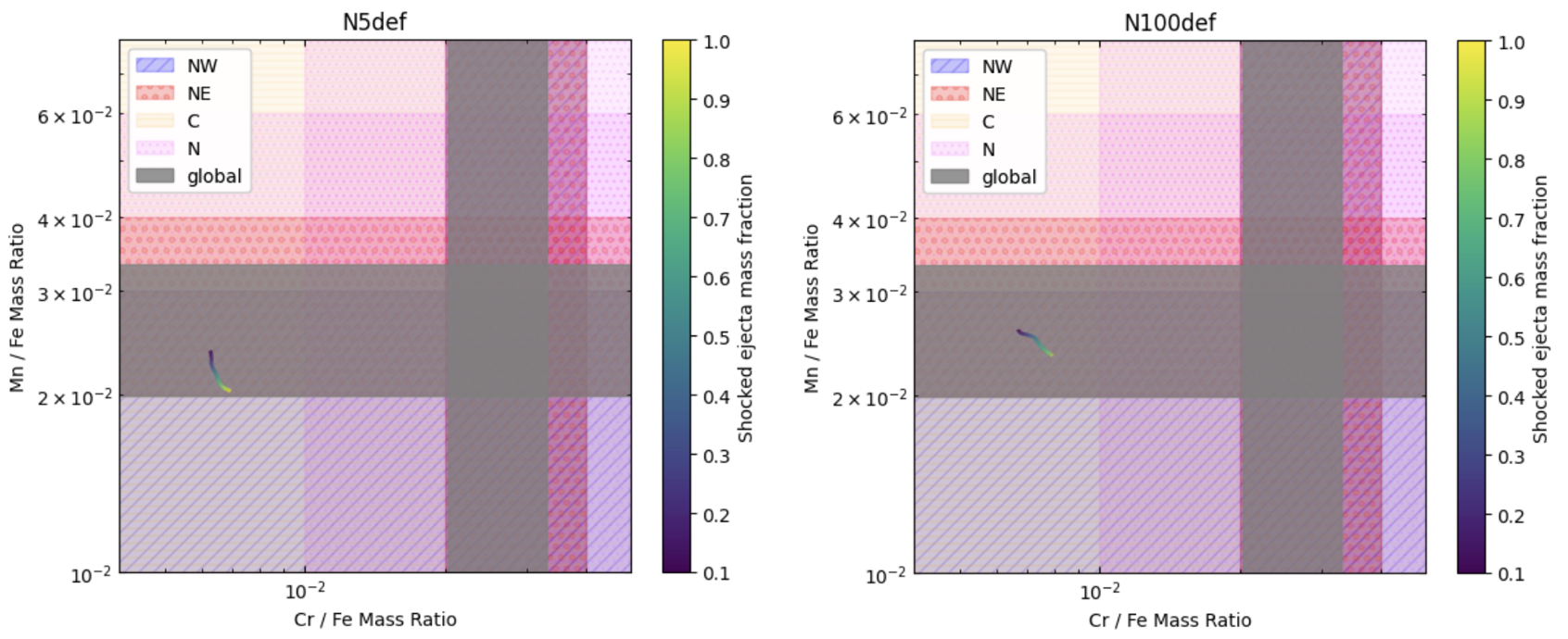}
    \caption{
    Comparison of the Cr/Fe and Mn/Fe mass ratios between observations and expectations from a Chandrasekhar mass WD exploding through 3D pure deflagration explosion mechanism from \cite{2014MNRAS.438.1762F} for a symmetric (100 ignition points, left panel) and asymmetric (5 ignition points, right panel) ignition of the explosion. Dashed stripes show the mass ratios from our study of Tycho's SNR with \textit{XRISM}/Resolve with a different color for each region and in grey for the simultaneous fit to all regions. Mass ratios from the model are shown as a color coded line as a function of the shocked ejecta mass fraction from 0 (black) to 1 (yellow) corresponding to all ejecta shocked. Mass ratios from the 3D models have been averaged over angles. The initial metallicity of the WD progenitor is solar and the initial central density of the WD is 2.9 $\times 10^9$ g cm$^{-3}$.}
    \label{fig:def}
    \end{figure*}

To provide constraints on the explosion mechanism and WD mass of Tycho's SNR, we compare the new \textit{XRISM}/Resolve Cr/Fe and Mn/Fe mass ratios with expectations from type Ia simulations of various supernova models. We compared with the same 1D models as in \cite{2025ApJ...986..216H} and with a set of additional models from 3D numerical simulations of supernovae. As most models produce radial stratification of the elements, we provide the mass ratios as a function of the mass fraction of shocked ejecta for 3D models, and for discrete values for 1D models. For the 3D models, the mass ratios arise from the average radial mass profiles of the ejecta elements. 
We show the mass ratios of Cr and Mn to Fe of Tycho's supernova remnant obtained in the NW, NE, C and N regions (individually and overall) compared to expectations from different supernova explosion models: pure deflagration (Fig. \ref{fig:def}), delayed detonation (Fig. \ref{fig:DDT}) and double detonation (Fig. \ref{fig:DD}) for a set of masses from near-Chandrasekhar to sub-Chandrasekhar exploding WDs.

In all figures showing these comparisons, our observational results are drawn as shaded stripes of different colors for each region of Tycho's SNR. The gray stripes show the average values obtained from  a simultaneous fit of all regions (see Sect. \ref{sec:analysis}).

\subsubsection{Deflagration}
\label{subsec:def}

We first compare in Figure \ref{fig:def} our results with a Chandrasekhar mass WD progenitor for two pure deflagration explosion mechanism models from \cite{2014MNRAS.438.1762F}: either with symmetric or asymmetric ignition (100 and 5 ignition kernels, respectively). In both cases, the single degenerate accreting WD has a central density of $2.9 \times 10^9$ g cm$^{-3}$ and an homogeneous mass composition of $^{12}$C (47.5$\%$), $^{16}$O (50 $\%$) and $^{22}$Ne (2.5 $\%$), corresponding approximately to a zero-age main-sequence progenitor with solar metallicity. The ejecta mass is 1.31 M$_{\odot}$ for the 100 ignition kernels (N100def) model while it is 0.372 M$_{\odot}$ for 5 ignition points (N5def). In those models, most of the Cr arises from the neutron-rich isotope $^{54}$Cr produced in the dense core by electron capture. Both models have been evolved to their supernova remnant phase in \cite{2021ApJ...906...93F}. The expected mass ratio is color coded as a function of the fraction of shocked ejecta mass, from black (10$\%$ shocked ejecta) to yellow (entirely shocked ejecta). 

As there are little radial variations of the ejecta elemental composition in pure deflagration models, the Cr/Fe and Mn/Fe mass ratios do not vary strongly with the fraction of the shocked ejecta mass (Fig. \ref{fig:MnCr}, where mass ratios from the deflagration models are shown as a function of the shocked ejecta mass fraction, and orange and red lines). While the average measured Mn/Fe mass ratio is in good agreement with the high Mn/Fe from both asymmetric and symmetric deflagration explosion models, the observed Cr/Fe mass ratio in Tycho's SNR ($>2 \times 10^{-2}$) is much higher than in the deflagration models ($<8 \times 10^{-3}$ for N100def and $<7 \times 10^{-3}$ for N5def).

On the basis of the inadequate Cr/Fe mass ratio, we confirm that deflagration models are not relevant for Tycho's normal type Ia SNR. The observed layered ejecta profile in Tycho's SNR, with Fe K emission located inner to the Si emission \citep{2001A&A...365L.218D}, is also not reproduced in deflagration models. This is consistent with the fact that deflagration models  do not reproduce the properties of normal type Ia (e.g., \citealt{2025A&ARv..33....1R}).

\subsubsection{Pure central detonation and delayed detonation}
\label{subsec:DDT}

    \begin{figure*}[h!]
    \includegraphics[scale=0.35]{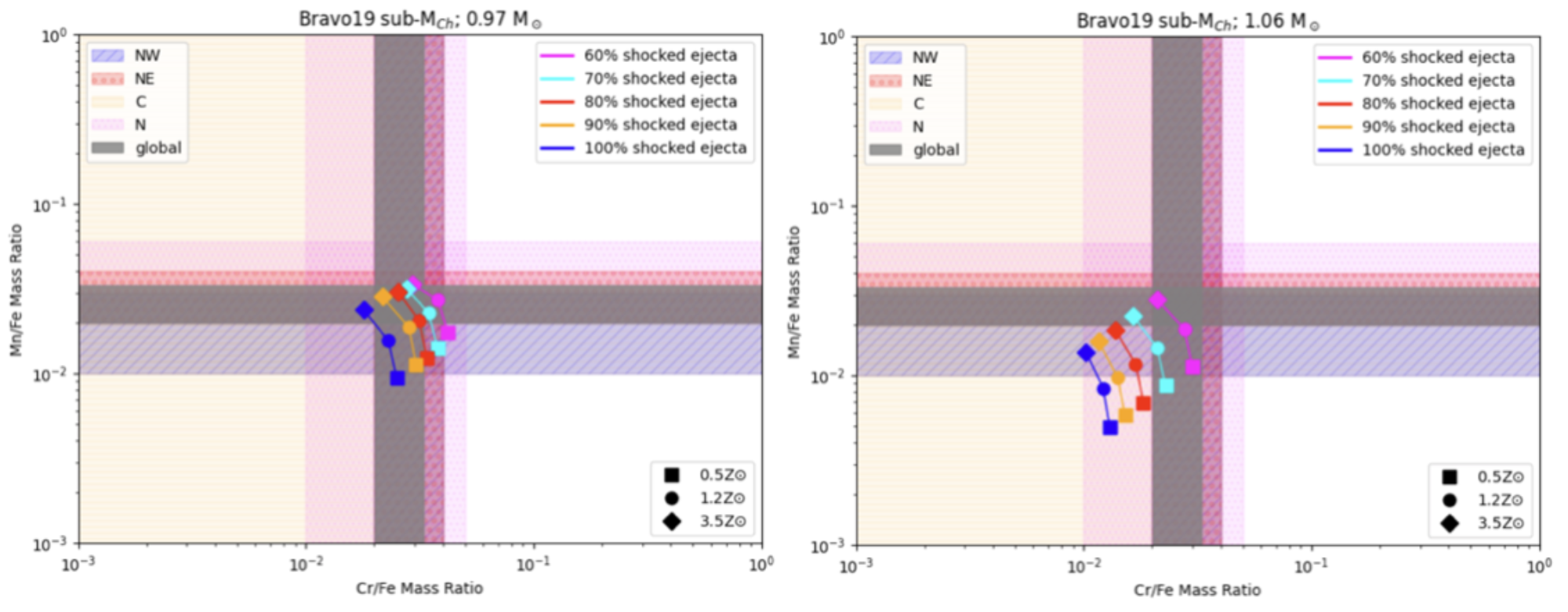} 
     \includegraphics[scale=0.35]{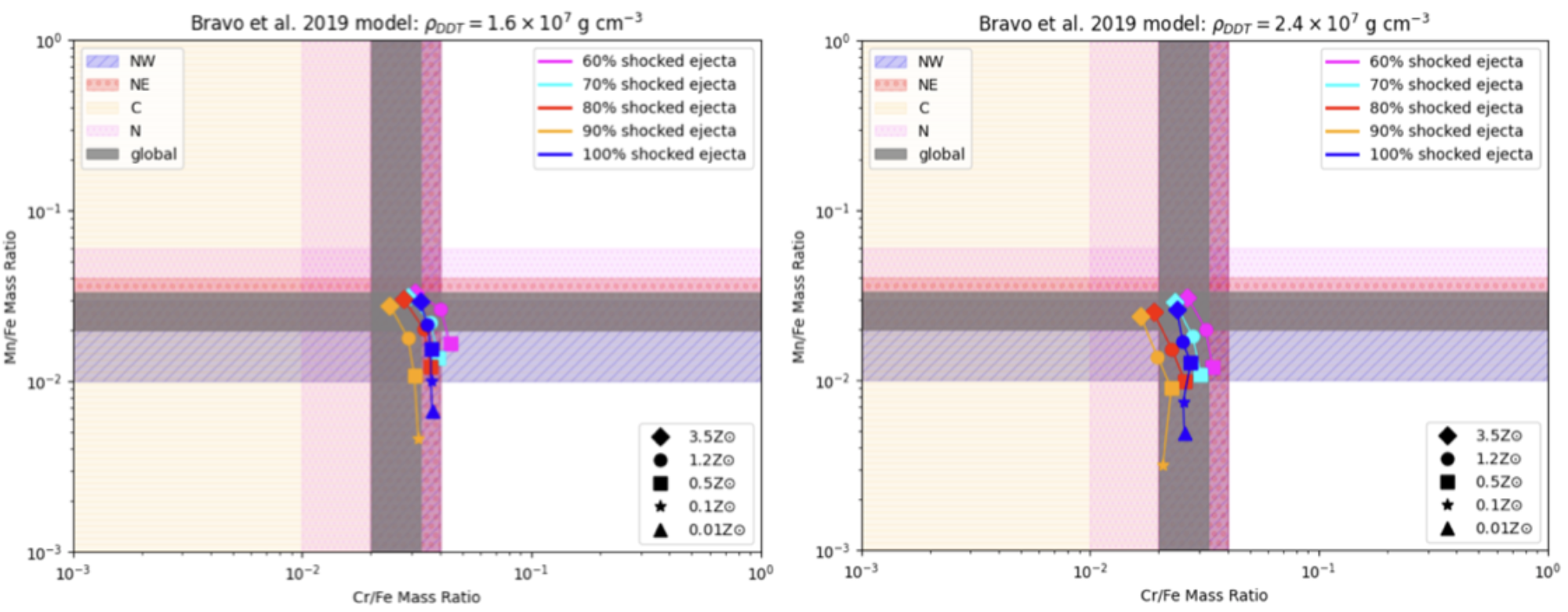} 
    \centering
    \includegraphics[scale=0.43]{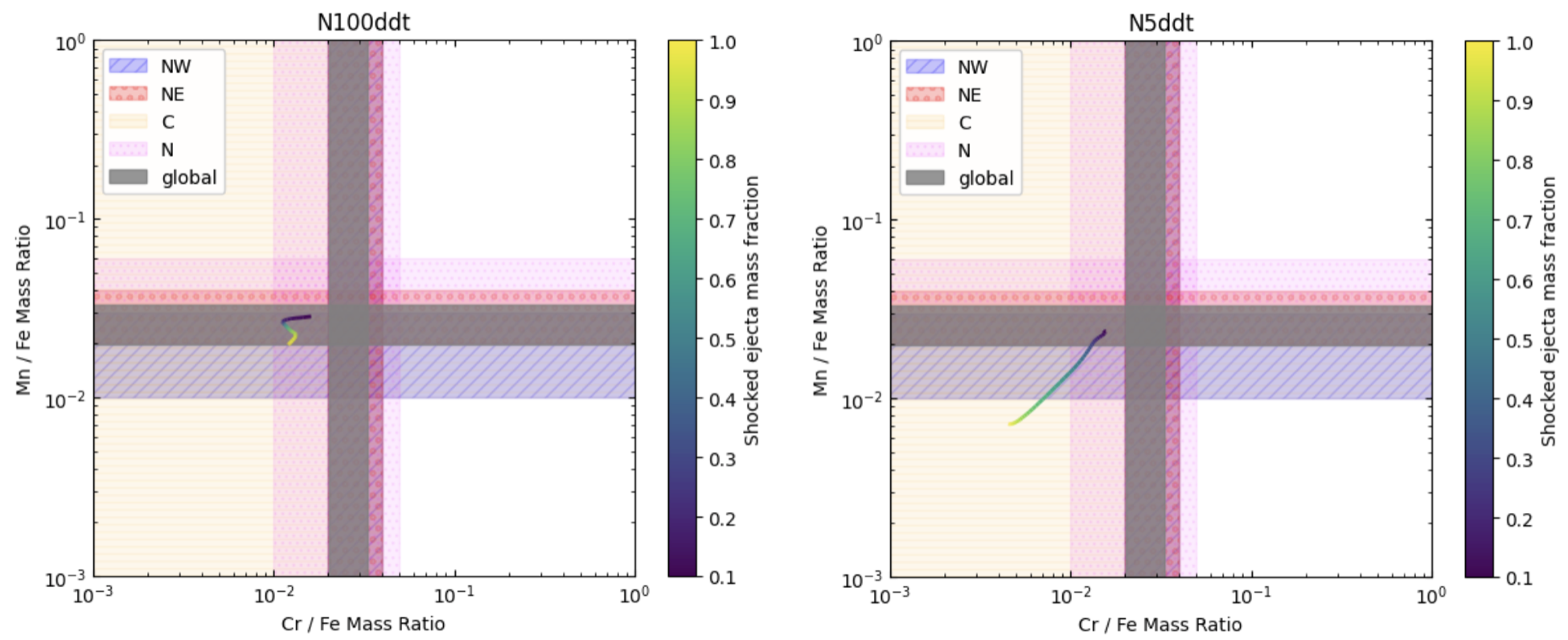}
    \caption{Comparison of the Cr/Fe and Mn/Fe mass ratios between observations and expectations from pure detonation and deflagration-to-detonation transition (DDT) models. Dashed stripes show the mass ratios from our study of Tycho's SNR with \textit{XRISM}/Resolve with a different color for each region and in grey for the simultaneous fit to all regions. Top panels: Comparison with 1D DDT models from \cite{2019MNRAS.482.4346B} with different initial metallicity for the WD progenitor and mass fraction of shocked ejecta. Left panels show sub-Chandrasekhar pure detonation cases and right panels Chandrasekhar mass with a WD density at the deflagration-to-detonation transition of $1.6\,10^7$ g cm$^{-3}$ (left) and $2.4\,10^7$ g cm$^{-3}$ (right). Bottom panels: Comparison with 3D DDT simulations from \cite{2013MNRAS.429.1156S} for a symmetric (100 ignition points, left panel) and asymmetric (5 ignition points, right panel) ignition of the explosion of a near-Chandrasekhar WD. The shocked ejecta mass fraction is shown as a color-coded line from 0 (black) to 1 (yellow) with all ejecta shocked. Mass ratios have been averaged over angles. }
    \label{fig:DDT}
    \end{figure*}
    
We then compare in Figure \ref{fig:DDT} our results with sub-Chandrasekhar mass detonation and Chandrasekhar mass deflagration-to-detonation transition models (DDT, or delayed detonation), considered as the most accepted explosion mechanism for normal type Ia supernovae \citep{2025A&ARv..33....1R}. The top panel shows 1D models, while the bottom panel has 3D models. The 1D models are from \cite{2019MNRAS.482.4346B}, following \cite{2025ApJ...986..216H} to allow a comparison with their findings based on the global mass ratios between IME and IGE. Four 1D models are shown for near and sub Chandrasekhar mass WD progenitors of different initial metallicity (Z) and WD density at deflagration to detonation transition. The Chandrasekhar models have a central density of $3 \times 10^9$ g cm$^{−3}$ and an explosion energy of about $1.3-1.4 \times 10^{51}$ erg s$^{-1}$, while the sub-Chandrasekhar models have a central density of $2.6-2.8 \times 10^7$ g cm$^{−3}$ and energy of about $1.1 \times 10^{51}$ erg s$^{-1}$ for 0.97 M$_{\odot}$ and $4.7-5.2 \times 10^7$ g cm$^{−3}$ and about $1.3 \times 10^{51}$ erg s$^{-1}$for 0.97 M$_{\odot}$. The different colors in the simulated mass ratios show different percentages of shocked ejecta mass. The 3D DDT models are from \cite{2013MNRAS.429.1156S} for a Chandrasekhar mass WD with a symmetric (N100ddt with 100 ignition points) and asymmetric (N5ddt with 5 ignition points) explosion ignition. The same initial conditions are assumed as for the deflagration models in terms of metallicity (solar) and central density of $2.9 \times 10^9$ g cm$^{−3}$. As for the deflagration case, both models have been evolved to their supernova remnant phase in \cite{2021ApJ...906...93F}. 

For all 1D models (top and middle panels of Fig. \ref{fig:DDT}), our results on Cr/Fe and Mn/Fe require a supersolar metallicity (3.5 Z$_{\odot}$) for the WD progenitor for a possible range of shocked mass fraction.

For 3D M$_{\rm Ch}$ DDT models (bottom panels of Fig. \ref{fig:DDT}), the Cr/Fe and Mn/Fe mass ratios are observed to vary with the shocked ejecta mass fraction, slightly and non linearly for the symmetric N100DDT, while it changes significantly and linearly for the symmetric N5DDT model from above 10$^{-2}$ down to 7$\times 10^{-3}$. To help understanding these behaviors, the mass ratios versus shocked mass fraction are shown with blue and green lines in Fig. \ref{fig:MnCr}. We find that the average Mn/Fe mass ratio of Tycho's SNR can be well reproduced for the symmetric N100DDT explosion (bottom left panel, Fig. \ref{fig:DDT}), but not Cr/Fe mass ratio, which is slightly too low. By extrapolating from 1D models, this indicates that a lower $\rho_{\rm{DDT}}$ is required. For the asymmetric N5DDT model (bottom right panel, Fig. \ref{fig:DDT}), the Mn/Fe mass ratio can also be reproduced but only for an unrealistically small shocked ejecta mass fraction ($<20 \%$), while the Cr/Fe mass ratio is too low, even for a small shocked mass fraction. As a realistic shocked ejecta mass fraction is estimated to be above 50$\%$ in Tycho's SNR \citep{2010ApJ...709.1387K}, both mass ratios become too low by a factor of a few. Decreasing $\rho_{\rm{DDT}}$ for the symmetric model might help to increase the Cr/Fe mass ratio, while a higher metallicity will increase the Mn/Fe mass ratio.

\subsubsection{Double detonation}
\label{subsec:DD}

  \begin{figure*}[h!]
    \centering
    \includegraphics[scale=0.35]{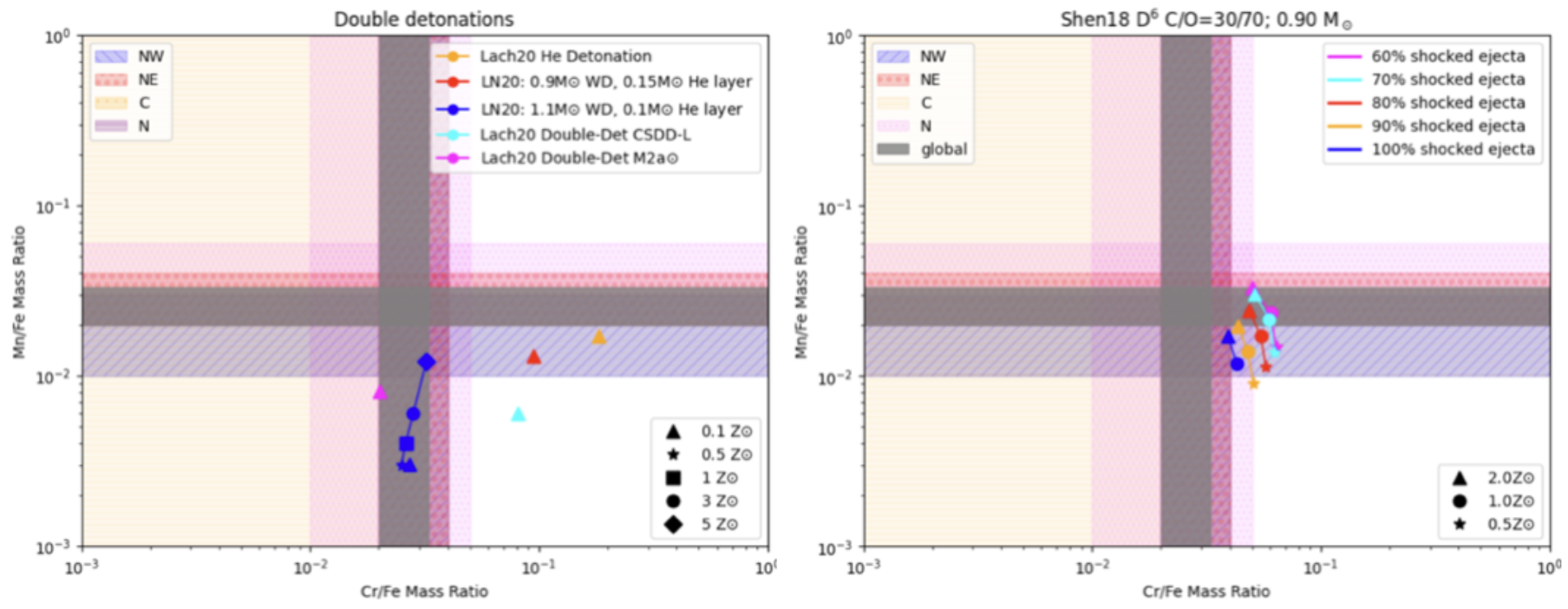} 
    \includegraphics[scale=0.35]{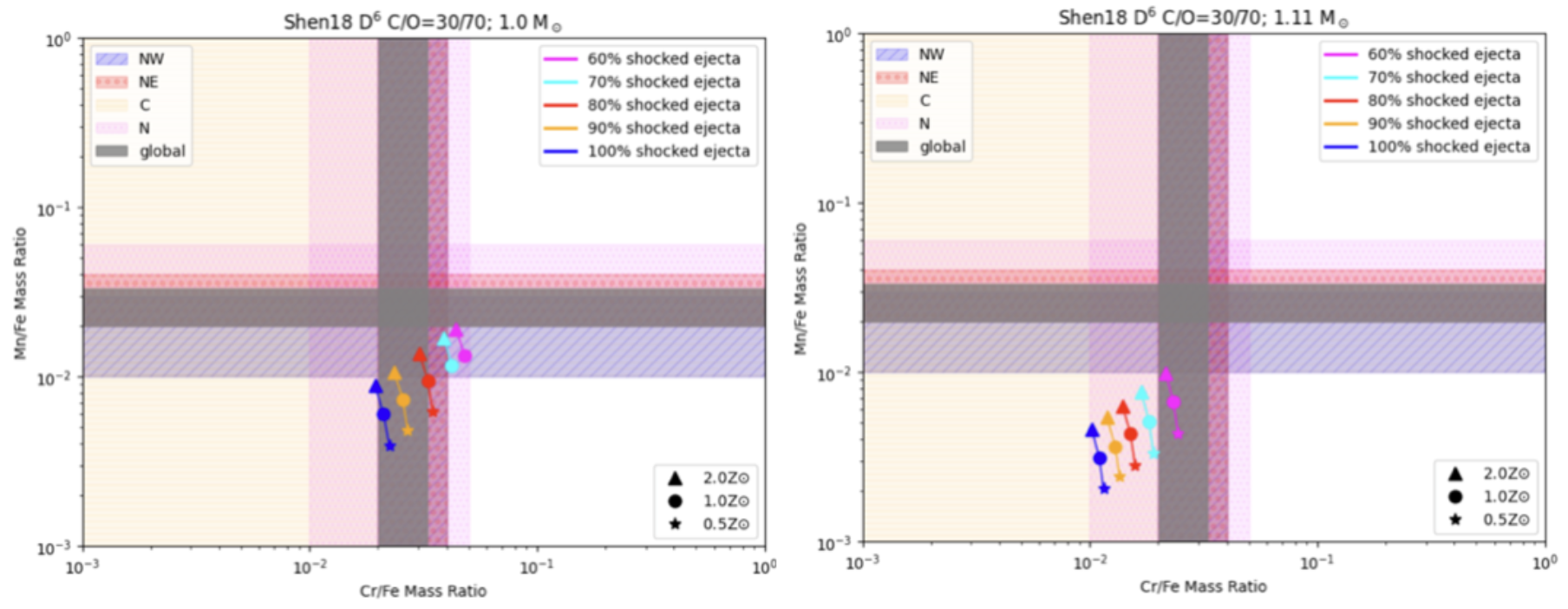} 
    \includegraphics[scale=0.5]{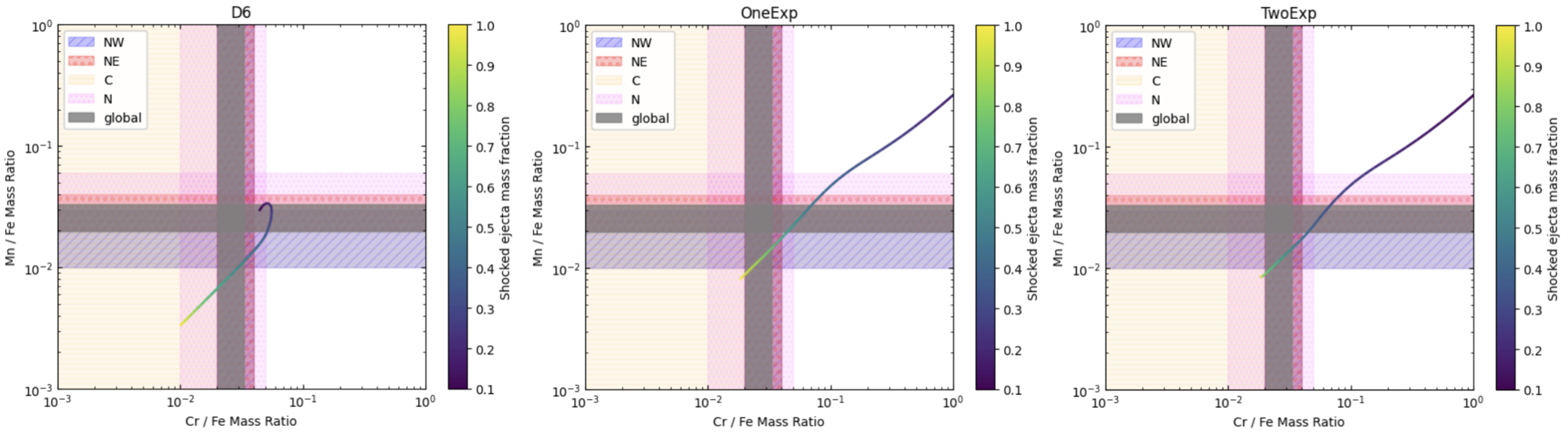}
    \caption{Comparison of the Cr/Fe and Mn/Fe mass ratios between observations and expectations from double detonation (DD) models of stable and unstable accretion (dynamically driven double-degenerate DD; D6). Dashed stripes show the mass ratios from our study of Tycho's SNR with \textit{XRISM}/Resolve with a different color for each region and in grey for the simultaneous fit to all regions. Top panel: comparison with 1D models for different WD masses, initial metallicities of the WD progenitor and mass fractions of shocked ejecta, as used in \cite{2025ApJ...986..216H}. Models with stable accretion are from \cite{2020A&A...644A.118L} and \cite{2020ApJ...888...80L} (left panel), and with unstable accretion from \cite{2018ApJ...854...52S} (right panels). Bottom panel: 3D D$^6$ models from \cite{2018MNRAS.475L..67T} (left panel) and from \cite{2022MNRAS.517.5260P} with the explosion of the primary WD (middle) and both primary and secondary WDs (right panel). The shocked ejecta mass fraction is shown as a color-coded line from 0 (black) to 1 (yellow) with all ejecta shocked. Mass ratios have been averaged over angles.}
    \label{fig:DD}
    \end{figure*}

We finally compare our results with double detonation models from stable and unstable accretion, which appear as a promising scenario for normal type Ia supernovae \citep{2025A&ARv..33....1R}. Figure \ref{fig:DD} presents the comparison of Tycho's Cr/Fe and Mn/Fe mass ratios with expectations from a set of double detonation models. Top left panel shows the comparison with stable accretion 2D models from \citet{2020A&A...644A.118L} and \citet{2020ApJ...888...80L}. All the other panels present unstable accretion from dynamically-driven double-degenerate double detonation (D$^6$) models. The three top right panels show the comparison with 1D models for 0.9, 1.0 and 1.11 M$_{\odot}$ from \citet{2018ApJ...854...52S}. The  bottom panels show 3D models with a single exploding WD (left and middle) and double exploding WDs (right). The left panel is for a 1 M$_{\odot}$ WD composed of a 50$\%$ carbon and 50$\%$ oxygen core, with an helium-dominated layer of 0.05 M$_{\odot}$ and a companion WD of 0.6M $_{\odot}$ \citep{2018MNRAS.475L..67T}, hereafter labelled "D6".
We also explore two scenarios from \citet{2022MNRAS.517.5260P}, where either only the primary WD explodes (bottom-middle panel), labelled "OneExp", or both WDs explode (bottom-right panel), labelled "TwoExp". The primary WD has a mass of 1.05 M$_{\odot}$, a central density of $4.8 \times 10^7$ g cm$^{-3}$, and is composed of a 50$\%$ carbon and 50$\%$ oxygen core (with solar metallicity), with a pure helium shell of 0.03 M$_{\odot}$. The companion WD has a mass of 0.7 M $_{\odot}$, a central density of $6.3 \times 10^6$ g cm$^{-3}$, with the same composition as the primary WD. These 3D models have been evolved to their supernova remnant phase in \cite{2022ApJ...930...92F} and \cite{2025ApJ...995...85F}, respectively. 

None of the 2D stable accretion models (left top panel in Fig. \ref{fig:DD}) reproduce the Mn/Fe mass ratio observed in Tycho's SNR, which is higher. The Cr/Fe mass ratio is reproduced for a 1.1 M$_{\odot}$ WD. The 1D unstable accretion models (right top and middle panels in Fig. \ref{fig:DD}) account for the Cr/Fe mass ratio for a 1 M$_{\odot}$ WD, but again with a too low Mn/Fe mass ratio. However, if we extrapolate linearly the model to a metallicity of above 3 solar, both Cr/Fe and Mn/Fe should be reproduced, providing that 80$\%$ of the ejecta mass is shocked.

For 3D models of unstable accretion (bottom panels in Fig. \ref{fig:DD}), a limited set of D$^6$ configurations has currently been simulated in 3D, assuming 1 M$_{\odot}$ WD mass and solar metallicity for the WD progenitor. The mass ratios of these models are shown as a function of the shocked ejecta mass fraction in Fig. \ref{fig:MnCr}. The explosion of the secondary WD does not change the mass ratios, as the Fe group elements are mainly produced by the explosion of the primary WD. As far as this diagnostic is concerned, the fate of the companion WD does not matter, it allows for a smaller fraction of shocked ejecta mass. The Cr/Fe mass ratio of Tycho's SNR is reproduced for a reasonable range of shocked ejecta mass fraction. However, the associated Mn/Fe mass ratio falls below that of Tycho's SNR or would require a lower shocked ejecta mass fraction. A higher metallicity for the WD progenitor should increase the Mn/Fe mass ratio while keeping a consistent Cr/Fe mass ratio, as illustrated for 1D D$^6$ models of \citet{2018ApJ...854...52S}.

\subsection{Neutron excess as tracer of the progenitor system}
\label{sec:disc_met}

\cite{2008ApJ...680L..33B} proposed a method to measure the neutron excess of progenitor system of type Ia SNe, based on Mn and Cr emission lines. They showed that the Mn/Cr mass ratio is strongly correlated with the zero-age main-sequence progenitor metallicity Z ($M_{\rm{Mn}}/M_{\rm{Cr}} = 5.3 \times Z^{0.65}$), as set by the neutron excess of the white dwarf material prior to thermonuclear runaway. From the Mn/Cr mass ratio derived from \textit{Suzaku} observations of Tycho's SNR \citep{2009PASJ...61S.167T}, they obtained $Z = 0.048^{+0.051}_{-0.036} = (3 \pm 2)$ $Z_\odot$, using \cite{2005ASPC..336...25A} as reference for solar metallicity. These results were interpreted as Tycho's SNR coming from the 'prompt' channel of Type Ia SNe.

More recently, \cite{2025ApJ...986..216H} have derived a metallicity of $Z \approx 0.019 \pm 0.007 \approx (1 \pm 0.37)\, Z_\odot$, based on their Mn/Cr mass ratio of $0.41 \pm 0.14$. They argued that the solar metallicity value better matches with Tycho's location \citep{2004A&A...418..989N} and allows for an older progenitor. 

\citet{2019MNRAS.482.4346B} have shown that increasing the metallicity leads to larger yields of the odd-atomic elements, such as Mn, but leaves most of the even-atomic elements, such as Cr, unaffected. The higher Mn flux from \textit{XRISM}/Resolve observations of Tycho's SNR provides a higher Mn/Cr mass ratio than in \cite{2025ApJ...986..216H}. With  \textit{XRISM}/Resolve results, we obtain a mean Mn/Cr mass ratio of $1.0 \pm 0.3$ (see sect. \ref{sec:mass}), which translates into a super-solar metallicity of $0.06 \pm 0.03 = (4 \pm 2)\, Z_\odot$, as initially proposed in \cite{2008ApJ...680L..33B}. The type Ia progenitor of Tycho’s SNR has then to arise from a prompt channel, rather than from an older progenitor with a solar metallicity.

  \begin{figure*}[h!]
    \centering
    \includegraphics[scale=0.31]{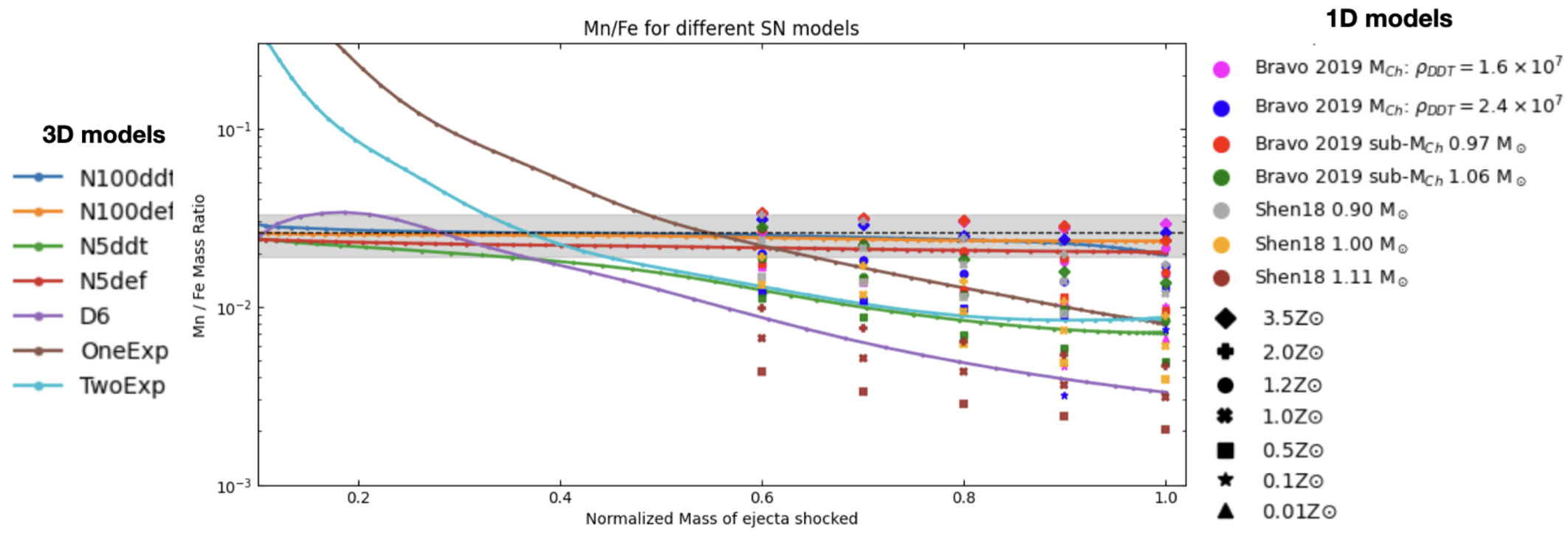}
    \includegraphics[scale=0.3]{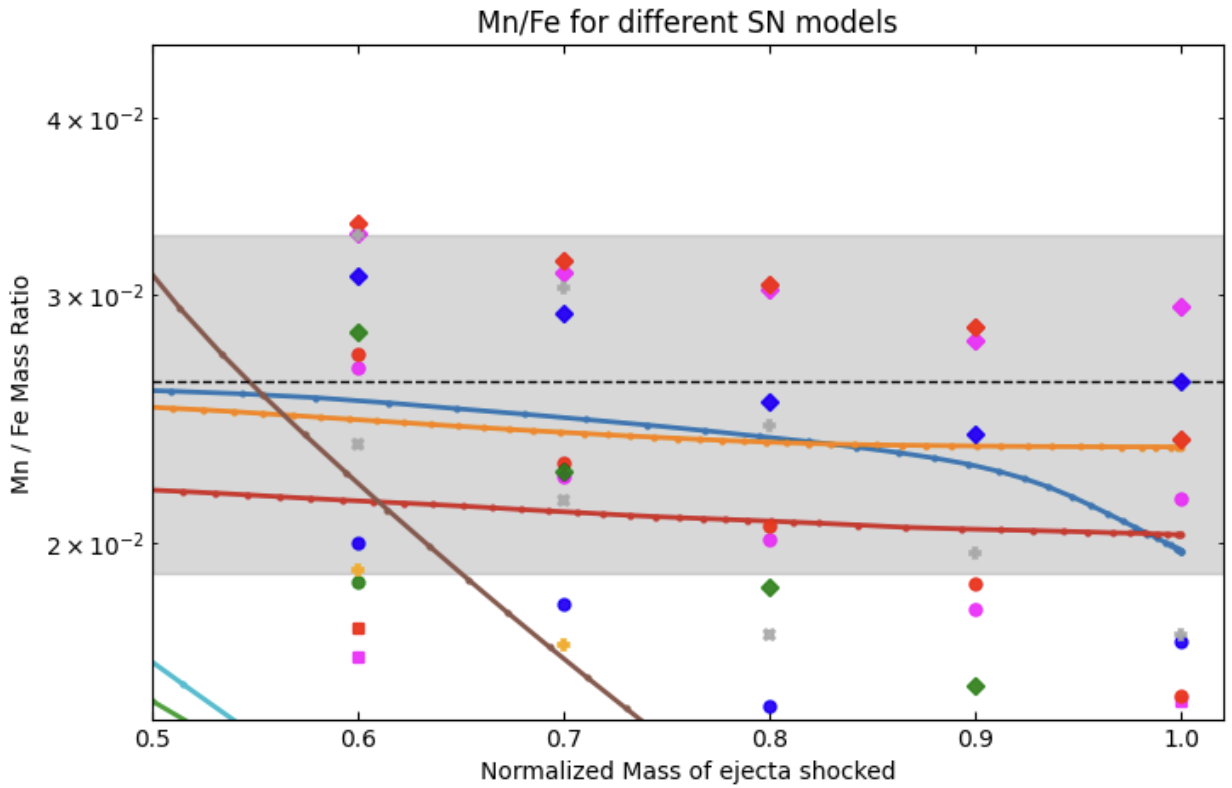}
    \includegraphics[scale=0.31]{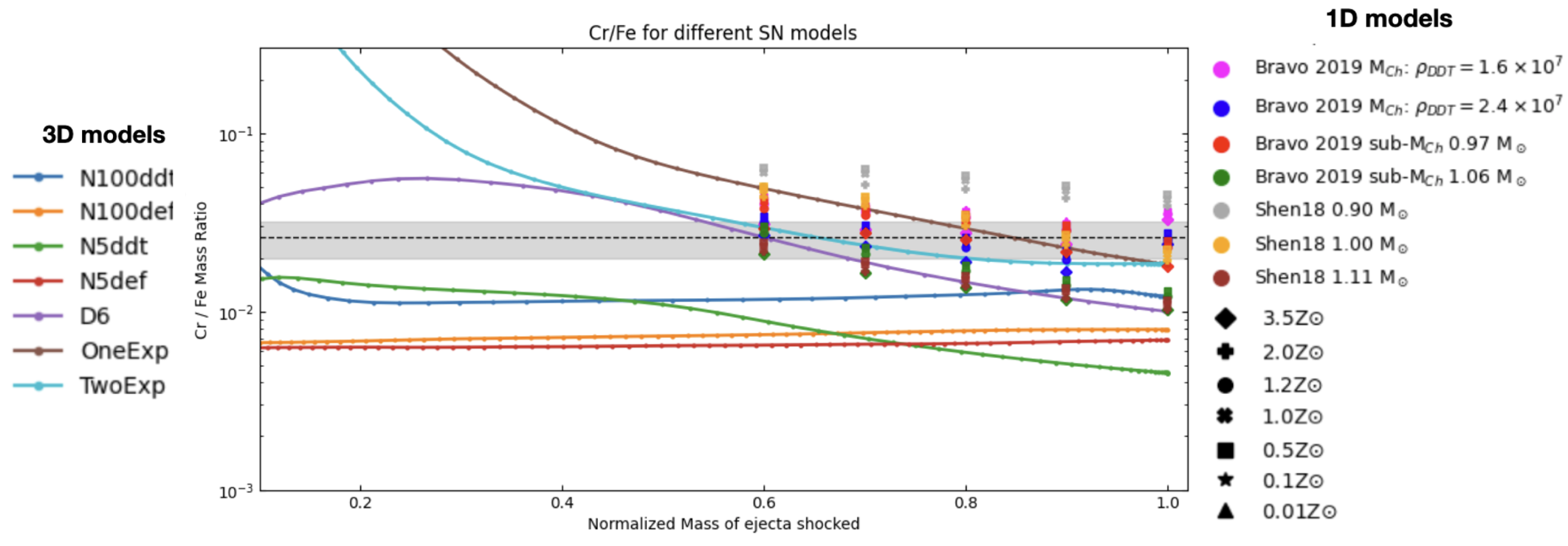}
    \includegraphics[scale=0.3]{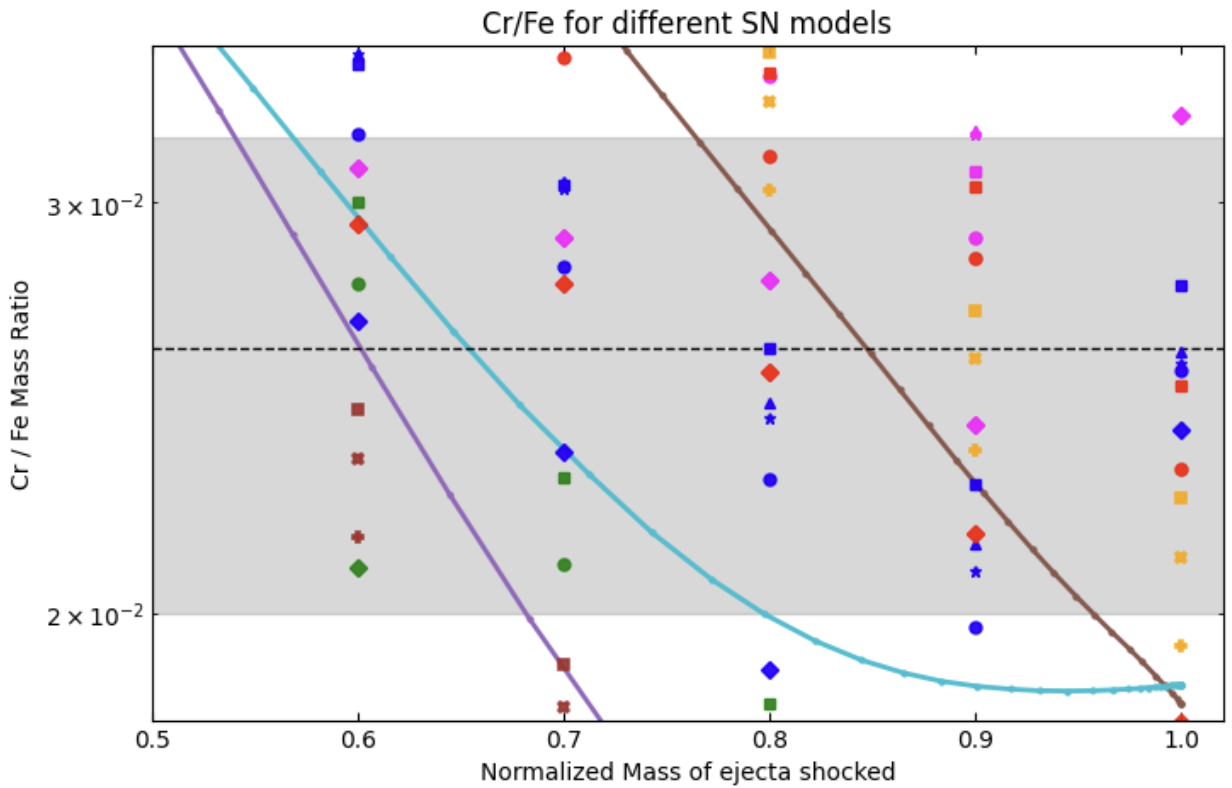}
    \includegraphics[scale=0.31]{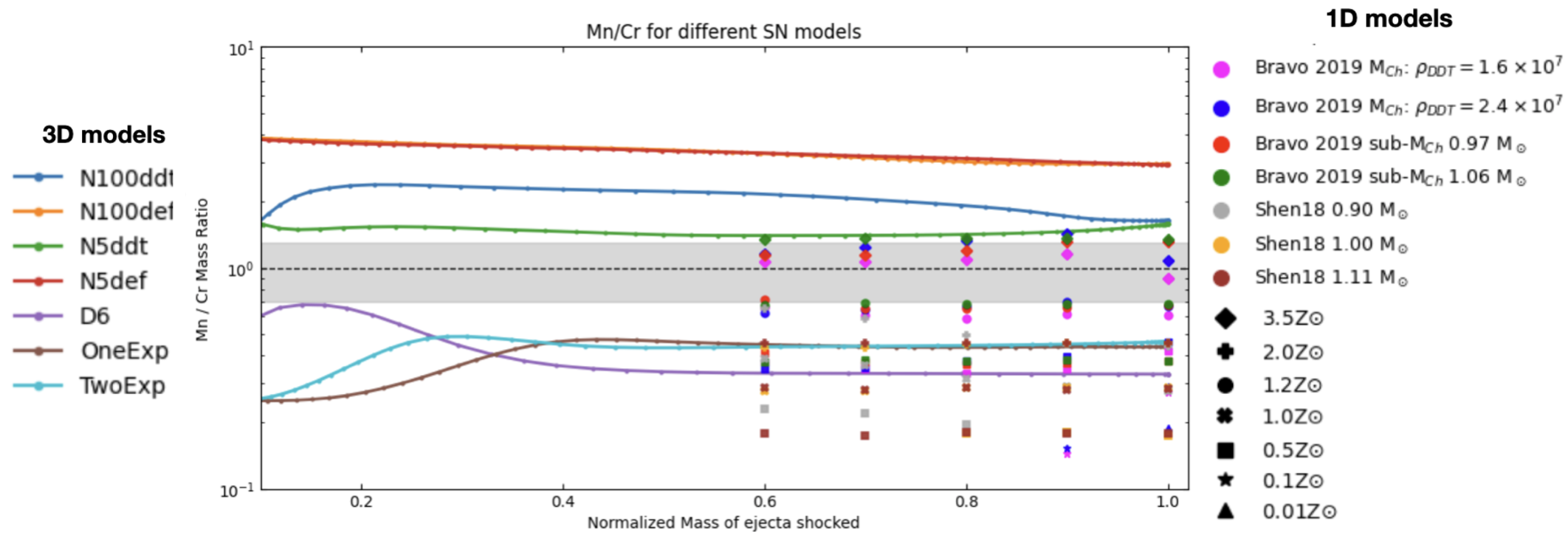}  
    \includegraphics[scale=0.3]{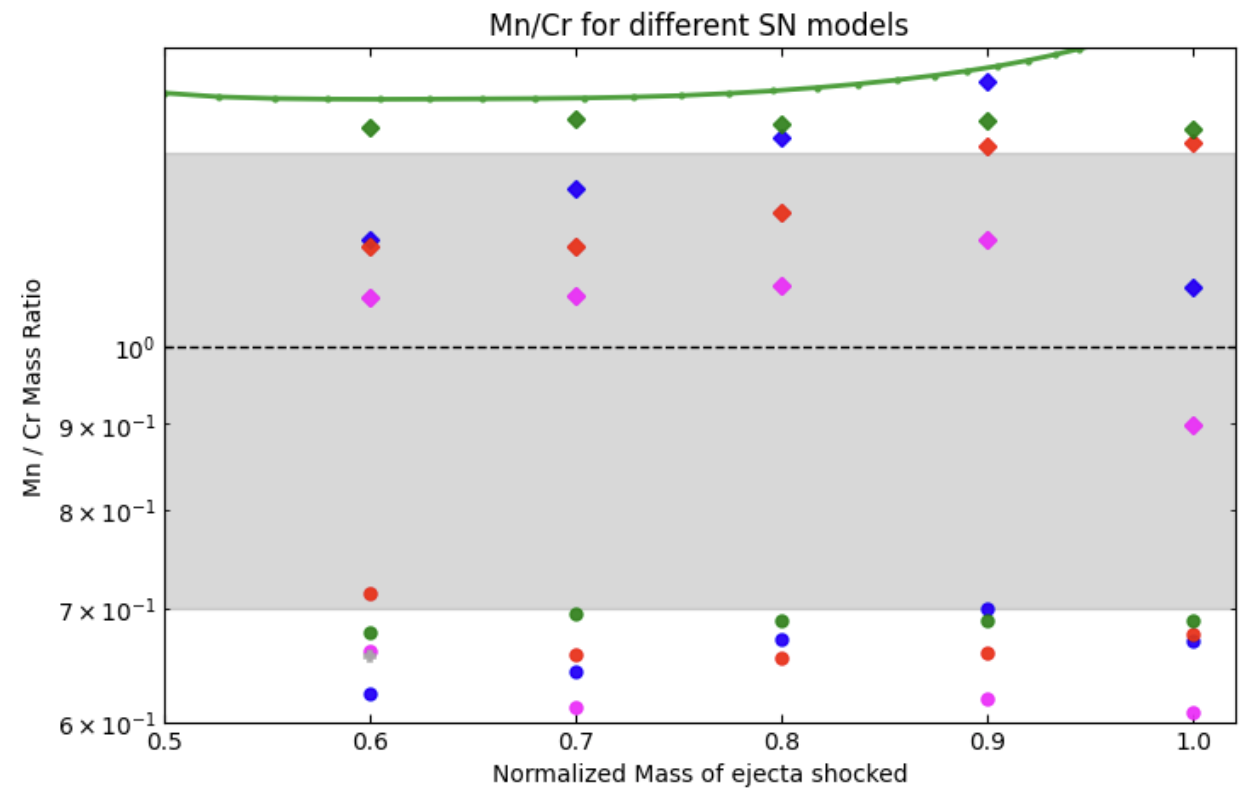} 
    \caption{Comparison of the Mn/Fe (top panel), Cr/Fe (middle panel) and Mn/Cr (bottom panel) mass ratios between observations and expectations from the SN models presented in Sect. \ref{sec:disc_Mch}. Left panels show a complete view of the comparison between observational data and simulations, right panels show a close view centered at the observational results. In all panels the gray stripes show the observational results for the global analysis.}
    \label{fig:MnCr}
    \end{figure*}

The bottom panel of Fig. \ref{fig:MnCr} compares the Mn/Cr mass ratio of Tycho's SNR with the models discussed in sect. \ref{sec:disc_Mch}. There is overall a limited dependence on the shocked ejecta mass fraction for both 1D (shown above 60$\%$) and 3D models (shown over a range of 10-100$\%$). However, for some models, there is some dependence. For the 1D D$^6$ 0.9 M$_\odot$ model, $M_{\rm{Mn}}/M_{\rm{Cr}}$ decreases with increasing shocked mass fraction. For Mch  (1D and 3D N100) DDT models, $M_{\rm{Mn}}/M_{\rm{Cr}}$ decreases when reaching the innermost ejecta (shocked ejecta mass fraction larger than 90$\%$). For the 3D (OneExp, TwoExp) models, $M_{\rm{Mn}}/M_{\rm{Cr}}$ decreases in the 40$\%$ outermost ejecta while it increases for \cite{2018MNRAS.475L..67T} D$^6$ model.

Tycho's Mn/Cr mass ratio is reproduced by near-Chandrasekhar DDT and sub-Chandrasekhar detonation 1D models with an oversolar metallicity (between 1.2 and 3.3 Z$_\odot$) for all shocked mass fraction between 60 and 100$\%$. 
The 1D D$^6$ models from \cite{2018ApJ...854...52S} require a low WD mass of 0.9 M$_\odot$, less than 60$\%$ of shocked ejecta mass fraction, and above 2 Z$_\odot$ metallicity.
3D (N100def and N5def) deflagration models provide a too high Mn/Cr mass ratio, driven by a too low Cr/Fe mass ratio. 3D DDT models get closer to Tycho's observation. The (solar metallicity) N5DDT is close to compatible with Tycho's Mn/Cr mass ratio with similar values to the 1D sub-Chandrasekhar (1.06 M$_\odot$) 3.5 Z$_\odot$ metallicity detonation model of \cite{2019MNRAS.482.4346B}. 
3D D$^6$ sub-Chandrasekhar models (1 M$_\odot$ WD D6, 1.05 M$_\odot$ WD OneExp and TwoExp) provide a too low Mn/Cr mass ratio, and require a higher than solar metallicity and/or a lower WD mass (as shown by the 1D D$^6$ models) to account for Tycho's mass ratio.

A high zero-age main sequence metallicity of Tycho's WD is required to reproduce the measured Mn/Cr mass ratio with all 1D (DDT and DD) models (1.2-3.5 Z$_\odot$), and is also suggested for 3D DDT (> 1 Z$_\odot$) and DD models. Such a high metallicity at Tycho's location is not favored. The progenitor metallicity (usually defined as the amount of $^{22}$Ne) introduces a neutron excess \citep{2003ApJ...590L..83T,2010ApJ...711L..66B,2016ApJ...818L..19M}. The neutron excess ($\eta = 1 - 2 Y_e$, where $Y_e$ is the electron fraction) is expected to be zero in WD composed only by $^{12}$C and $^{16}$O. For sub-Chandrasekhar WDs, the metallicity is the main source of neutronisation. As shown in Fig. \ref{fig:MnCr} (bottom panel) for 1D D$^6$ models of \cite{2018ApJ...854...52S}, decreasing the mass of the progenitor WD helps for the neutronisation with a lower requirement on the metallicity.

For near-Chandrasekhar mass WD, the metallicity is not the only factor. $\eta$ can increase as a result of weak interactions occurring during various evolutionary stages of the progenitor. Neutronisation occurs during the carbon simmering, which is a long pre-explosive phase during which slow central carbon burning drives convection and gradually heats the core until nuclear energy production overcomes neutrino cooling and triggers runaway ignition \citep{2004ApJ...607..921W,2004ApJ...616.1102W,2008ApJ...678.1158P}. During this phase, electron capture on the product of carbon burning increases the neutron excess $\eta$ \citep{2008ApJ...677..160C,2008ApJ...673.1009P,2016ApJ...825...57M}. In addition, in the neutron-rich Nuclear Statistical Equilibrium (n-NSE) phase, the innermost~0.2 solar masses reach densities high enough for electron capture to occur during nucleosynthesis. This is however not a viable source of neutron excess for Tycho's SNR, as the reverse shock has not reached the innermost ejecta. Both mechanisms drive the nuclear statistical equilibrium away from producing mainly $^{56}$Ni and instead favors more neutron-rich isotopes such as $^{55}$Mn and $^{58}$Ni \citep{1999ApJS..125..439I}. As shown in Fig. \ref{fig:MnCr} (bottom right panel), a lower deflagration-to-detonation transition density helps in reducing the metallicity requirement in 1D DDT models.

\subsection{Summary}
\label{sec:sum}
From the comparison of the Cr/Fe, Mn/Fe and Mn/Cr mass ratios, from the \textit{XRISM}/Resolve study of Tycho's SNR with expectations from various models in Fig. \ref{fig:MnCr}, we restrict the possible progenitors and models for Tycho's SNR. Within the considered models, the 3D pure deflagration models are not consistent with our measurements.

1D pure sub-M$_{\rm Ch}$ detonation and M$_{\rm Ch}$ delayed detonation models can reproduce the three mass ratios only for a super solar metallicity WD progenitor (3.5 Z$_\odot$). For the sub-Chandrasekhar WD mass of 1.06 M$_{\odot}$, 60$\%$ maximum of shocked ejecta mass is allowed (a metallicity down to 1.2 Z$_\odot$ is acceptable). For a slightly lower WD mass of 0.97 M$_{\odot}$, the possible shocked ejecta mass fraction is 60-100 $\%$. For Chandrasekhar mass WDs, we find that a 60-100$\%$ shocked fraction is required for a density at deflagration-to-detonation transition of $1.6\times 10^7$ g cm$^{-3}$, while it reduces to 60-70 $\%$ for $\rho_{\rm{DDT}} = 2.4 \times 10^7$ g cm$^{-3}$.

Previous works estimate around 75\% to 95\% of shocked ejecta mass in Tycho's SNR \citep{2015ApJ...808...49K,2025ApJ...986..216H}, in agreement with our results for 1D delayed detonation models for 0.97 M$_{\odot}$ exploding WD and for deflagration-to-detonation transition at density at $1.6\times 10^7$ g cm$^{-3}$ for Chandrasekhar mass WD. A lower shocked ejecta mass fraction is still allowed in our results for detonation with larger sub-Chandrasekhar WD mass of 1.06 M$_{\odot}$ and deflagration-to-detonation transition at larger density $\rho_{\rm{DDT}} = 2.4 \times 10^7$ g cm$^{-3}$ for Chandrasekhar mass WD.

Delayed detonation models have key physical processes, such as buoyancy, that are intrinsically 3D \citep{2025A&ARv..33....1R}. We have thus explored two 3D DDT simulations of solar metallicity Chandrasekhar mass (symmetric versus asymmetric). They produce a too low Cr/Fe mass ratio.

For the double detonation (dynamically-driven double-degenerate) mechanism, none of the presented models match Tycho's Cr/Fe, Mn/Fe and Mn/Cr mass ratios. Specifically Mn/Cr is too low in these models. However, only a limited set of D$^6$ configurations has been simulated in 3D. A lower WD mass and/or a higher metallicity should provide a better match to Tycho's properties, as suggested by 1D D$^6$ simulations. 

A recent example in favor of the double detonation sub-Chandrasekhar mass exploding WD scenario is SNR 0509-67.5, with the presence of Fe XIV \citep{PhysRevLett.123.041101}, a double ring of Ca XIV emission \citep{2025NatAs...9.1356D} and a He-rich environment \citep{2026MNRAS.548ag596D}.

To reproduce the mass ratios of Cr/Fe, Mn/Fe and Mn/Cr in Tycho's SNR, all 1D models require a high metallicity zero-age main-sequence progenitor, and favor a prompt channel, although such a metallicity is not favored in Tycho's location. A high metallicity progenitor is also found in the supernova remnant 3C 397 with \textit{XRISM}/Resolve \citep{2026PASJ..tmp..202N}. 
 An increase in metallicity results in a higher initial abundance of $^{22}$Ne, which is produced primarily from the CNO-cycle material. The initial $^{22}$Ne abundance decreases the electron fraction $Y_{\rm e}$, enhancing the neutron excess of the white dwarf and therefore the production of neutron-rich iron-group nuclei \citep{1986A&A...158...17T}. In sub-Mch WD, the central density is generally too low to reach nuclear statistical equilibrium and electron capture reactions, and IGE are produced during the $\alpha$-rich freeze-out. The enhanced $^{22}$Ne abundance lowers the electron fraction and the abundance of free protons leading to less $^{55}$Co destruction and even more stable $^{55}$Mn from its decay \citep{2021A&A...656A..94G}.
As discussed in Sect. \ref{sec:disc_met}, for near-Mch WDs, other mechanisms can increase the neutron excess, such as carbon simmering, which produces neutronized species like Ni and Mn. However, unlike in 3C397 \citep{2015ApJ...801L..31Y}, Ni is not significantly detected in Tycho's SNR with {\it XRISM}/Resolve. The mass ratio of Ni/Fe is found to be below $4\%$. The neutronization degree in the shocked ejecta of Tycho's SNR is thus moderate.

Near-Chandrasekhar mass DDT models produce a large Ni/Fe mass fraction, even for low progenitor metallicity \citep[see Fig. 4 in][]{2015ApJ...801L..31Y}. They are thus disfavored for Tycho's SNR, unless the innermost part of the ejecta has not been shocked yet and most of the detected Cr and Mn are synthesized farther out in the explosive Si burning regime. 
Sub-Chandrasekhar detonation models produce less Ni, especially when the white dwarf's mass and metallicity are low, as shown in Fig. 4 in \citealt{2015ApJ...801L..31Y}. Investigation of lower WD mass in 3D models are thus favored for Tycho's SNR.

To further constrain the progenitor mass and explosion mechanism, we thus need on one hand 3D models exploring a higher degree of neutronization with a lower sub-Chandrasekhar mass WD and possibly higher than solar zero-age main-sequence metallicity progenitor. On the other hand, future observations with \textit{NewAthena}/XIFU , thanks to its higher sensitivity and spatial resolution \citep{2025ExA....59...18P}, will provide stringent constraints on the faint iron-group emission lines of Ti, Cr, Mn, Ni  and their location in the ejecta with full spatially-resolved high-resolution spectroscopy of Tycho's SNR.

\section{Conclusions}
\label{sec:conclusions}
In this paper we investigated the spatial distribution and physical properties of Chromium and Manganese in Tycho’s SNR using high-resolution X-ray spectroscopy with \textit{XRISM}/Resolve. The study of these elements provides a powerful probe of the innermost layers of Type Ia supernovae and offers crucial constraints on explosion mechanisms and progenitor systems.
By analyzing \textit{XRISM}/Resolve observations acquired during the PV and AO1 phases, we were able to map the spatial distribution of Cr and Mn with unprecedented precision. We find that both Cr and Mn emission are predominantly localized in the bright Fe rim, spanning from northeast to northwest. No Cr or Mn emission line is significantly detected in the southern part, where the emission is less bright and exposure time limited. 

For each pointing, we determined the mass ratios Mn/Fe and Cr/Fe. The Cr/Fe ratios are consistent with previous measurements, while the Mn/Fe ratios are significantly higher than those inferred from earlier \textit{Suzaku} observations. 
The derived mass ratios are remarkably uniform within error bars across the observed regions, indicating a chemically homogeneous plasma within the emitting shocked ejecta. 
Comparisons with theoretical models of Type Ia supernova explosions reveal very good agreement with 1D delayed detonation models for Chandrasekhar-mass progenitors, as well as 1D pure-detonation models for sub-Chandrasekhar-mass progenitors with 1 M$_\odot$. In both cases, observational results require supersolar metallicity of the zero-age main-sequence progenitor (3.5Z$_\odot$). Adding the upper limit on the Ni/Fe mass ratio, a moderate neutronization of Tycho's ejecta is required, which points either to a near-Chandrasekhar mass DDT model, provided that Ni is mostly still unshocked, or to a sub-Chandrasekhar white dwarf detonation model. The explored sub-Chandrasekhar-mass double-detonation explosion mechanism models, from both 1D and 3D simulations, have too low progenitor metallicities (1-2 Z$_\odot$) to reproduce Tycho's Mn/Fe and Mn/Cr mass ratios. While these results do not uniquely determine the nature of the progenitor system, they significantly tighten the constraints on the exploding star mass, the explosion mechanism and especially on the neutronization level. They also point to the need of 3D models with lower WD mass and/or possibly high metallicity progenitors ($>$Z$_\odot$). Overall, this study demonstrates the transformative impact of \textit{XRISM}/Resolve on the study of faint Fe-group elements and highlights its potential for advancing our understanding of Type Ia supernova progenitors and explosion mechanisms.

The high-spectral resolution study of Fe-group elements, such as Cr, Mn and Ni faint emission lines, both in spatial distribution and physical properties, is a powerful approach to investigate the origin of type Ia supernovae in terms of WD mass and metallicity, and explosion mechanism. This will be one of the focus of the new X-ray observatory \textit{NewAthena} with spatially-resolved high-spectral resolution studies of supernova remnants with X-IFU instrument. With NewAthena's higher spatial resolution and effective area compared to XRISM telescopes, X-IFU will provide spectroscopic information at much smaller scale on the localization of Cr and Mn, and Ni, and their physical properties, allowing full comparison with 3D simulations of supernovae. 

\begin{acknowledgements}
RG, AD, LG, JB, FA, acknowledge financial support from the Centre national d’études spatiales (CNES), France (ROR:
https://ror.org/04h1h0y33) within the framework of the NewAthena and XRISM missions.

\end{acknowledgements}

\bibliographystyle{aa}
\bibliography{references}

\appendix
\begin{onecolumn}

\section{GMCA images}
\label{sec:app_GMCA}

Fig. \ref{fig:gmca} shows two flux images representing the Fe and synchrotron spatial distributions generated by applying the GMCA method \citep{2015ITSP...63.1199B} on \textit{Chandra} observations \citep{2023A&A...680A..80G}.

    \begin{figure}[ht!]
    \centering
    \includegraphics[scale=0.3]{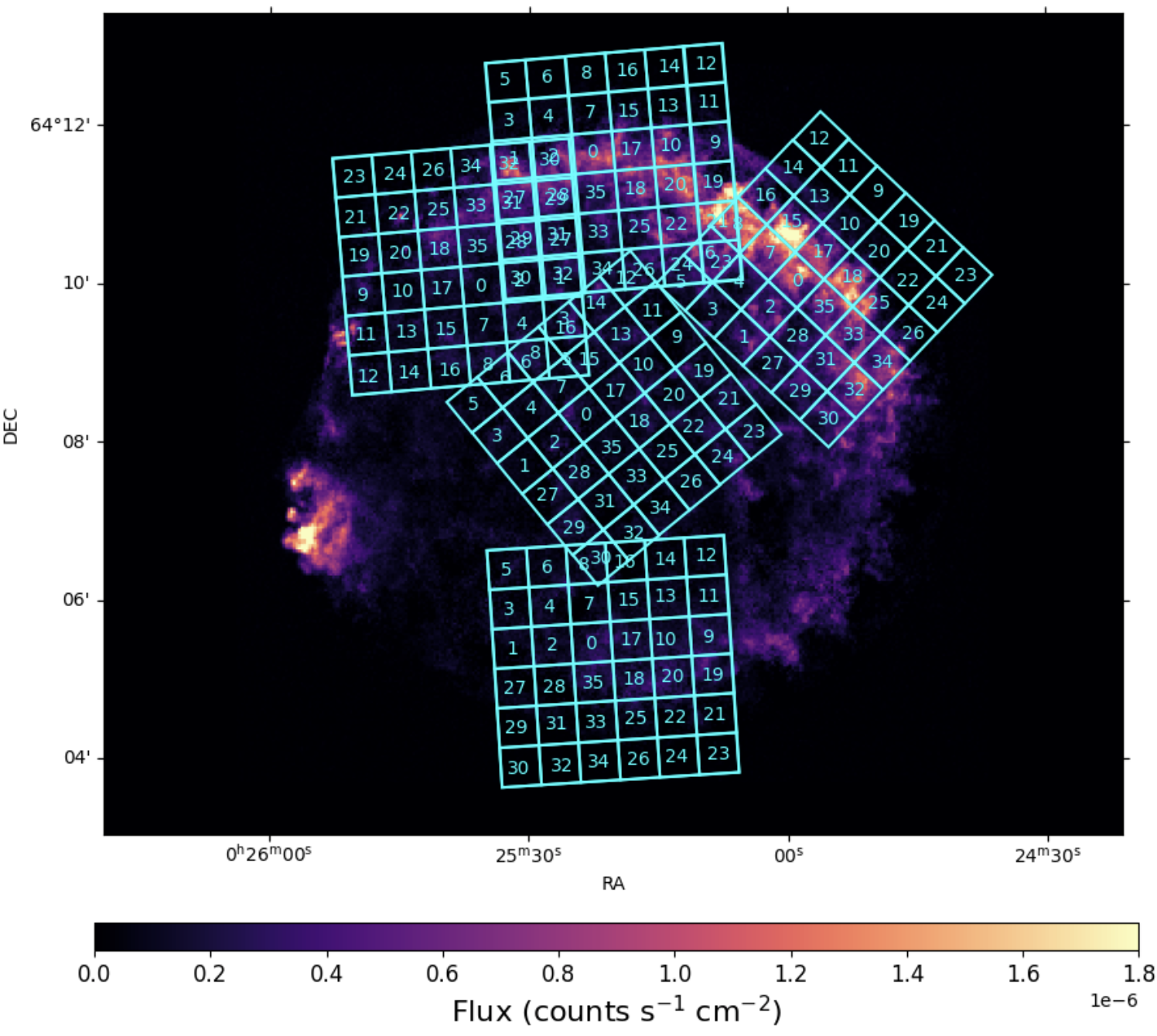}
    \includegraphics[scale=0.3]{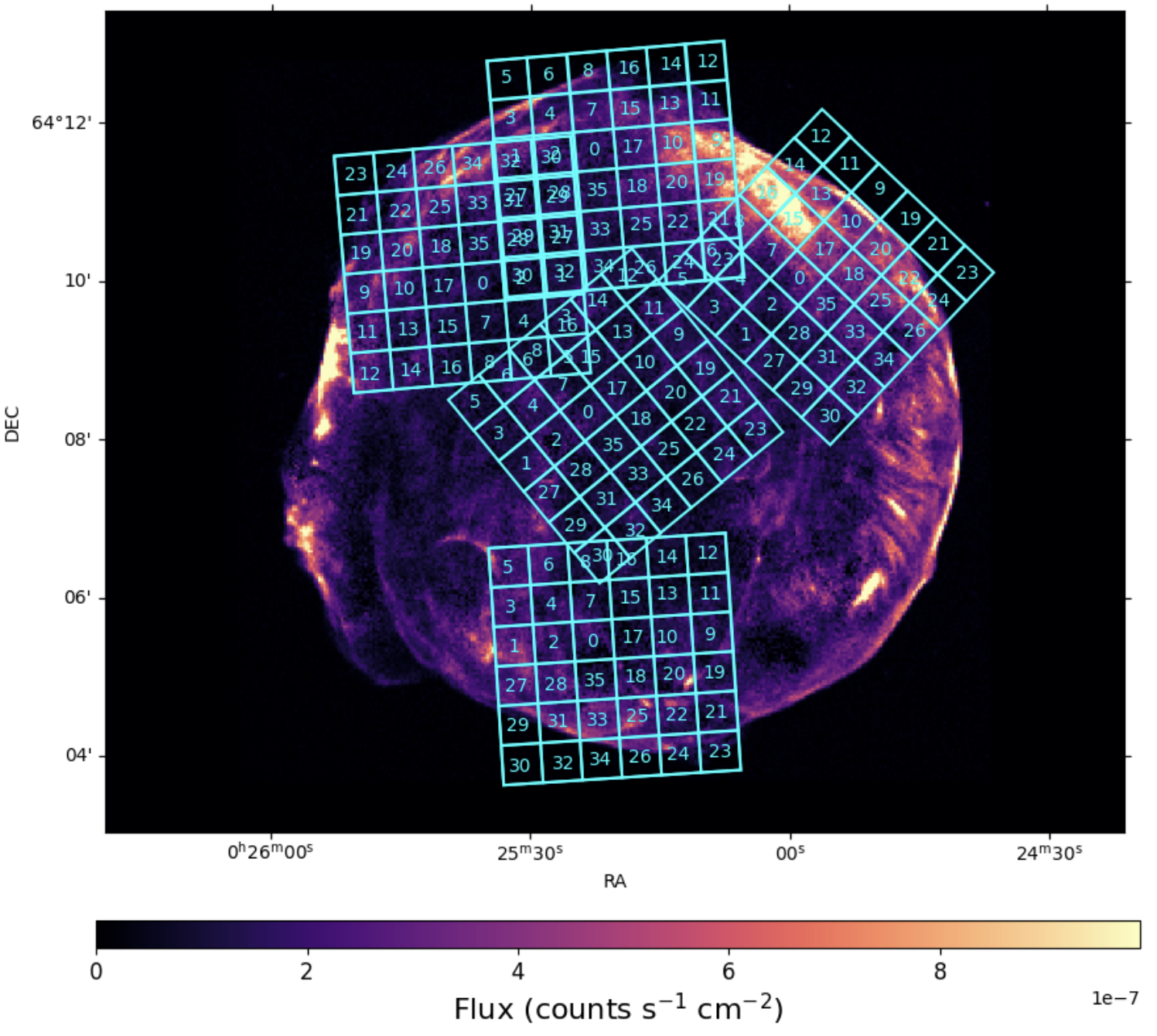}
    \caption{\textit{Left panel}: GMCA flux image of the Fe spatial distribution. \textit{Right panel}: GMCA flux image of the synchrotron spatial distribution.}
    \label{fig:gmca}
    \end{figure}

\section{Spectra and best fit parameters}
\label{sec:app_spectra}
Fig. \ref{fig:spec_NW_rows}, \ref{fig:spec_NE_rows}, \ref{fig:spec_N_rows} and \ref{fig:spec_C_squares} show the spectra of the sub-regions shown in Fig. \ref{fig:spatial} for the NW, NE, N and C pointings, together with their best fit model (Eq. \ref{eq:mod_gauss}). Fig. \ref{fig:spec_global_pshock} shows the four spectra for which we calculated the Cr/Fe and Mn/Fe mass ratios, together with their best fit model (Eq. \ref{eq:mod_pshock}. Best fit parameters are shown in Tab. \ref{tab:gauss}. Fig. \ref{fig:global_spec_average}) shows the simultaneous fit of all four spectra shown in Fig. \ref{fig:spec_global_pshock}, with the best fit model as discussed in Sect. \ref{sec:analysis}.

\begin{table*}[h]
\centering
        \caption{Best fit parameters of the Gaussian lines associated with Cr and Mn when fitting with Eq. \ref{eq:mod_gauss}. Errors are at 90\% confidence level. In the first column, the region subscripts indicate which rows of pixels are included in the spectrum. Row 0 is at the border of the remnant and row 5 is the innermost part of the pointing.} 
                \footnotesize
                \begin{tabular}{c c c c c c c c c c c }
                    \hline\hline
                    & & Cr & & & Mn & & & & Pow &\\
                    \hline
                    Reg & c$_1$ & Centroid  & N ($\times 10^{-6}$) & $\Delta C^{(a)}$ & Centroid  & N ($\times 10^{-6}$) & $\Delta C^{(a)}$ & c$_2$ & Index & N$^{(c)}$ ($\times10^{-2}$)\\
                    & & (keV) & (ph s$^{-1}$ cm$^{-2}$) &  & (keV) & (ph s$^{-1}$  cm$^{-2}$) & & & & (ph keV$^{-1}$ cm$^{-2}$)\\
                    \hline
                    \vspace{0.1cm}
                    NW$_{01}$ & 0.03 & $5.50^{+0.07}_{-0.05}$ & $0.9^{+0.8}_{-0.8}$ & 4 & $5.92^{+0.04}_{-0.04}$  & $1.2^{+0.8}_{-0.7}$ & 8 & 0.05 & $3.1^{+0.1}_{-0.1}$ & $0.9^{+0.2}_{-0.1}$ \\
                    \vspace{0.1cm}
                    NW$_{12}$ & 0.09 & $5.47^{+0.05}_{-0.06}$ & $1.6^{+1.1}_{-1.1}$ & 7 & $5.96^{+0.04}_{-0.04}$ & $1.8^{+1.0}_{-0.9}$ & 12 & 0.09 & $3.06^{+0.08}_{-0.08}$ & $1.1^{+0.2}_{-0.1}$\\
                    \vspace{0.1cm}
                    NW$_{23}$ & 0.14 & $5.46^{+0.03}_{-0.03}$ &  $2.4^{+1.1}_{-1.1}$ & 16 & $5.97^{+0.04}_{-0.03}$ & $2.0^{+0.9}_{-0.9}$ & 15 & 0.09 & $2.91^{+0.08}_{-0.08}$ & $0.9^{+0.1}_{-0.1}$ \\
                    \vspace{0.1cm}
                    NW$_{34}$& 0.13 &  $5.43^{+0.04}_{-0.03}$ & $1.1^{+0.8}_{-0.8}$ & 16 & $5.93^{+0.06}_{-0.06}$ & $0.6^{+0.5}_{-0.6}$ & 6 & 0.07 & $2.84^{+0.09}_{-0.09}$ & $0.62^{+0.10}_{-0.09}$\\
                    \vspace{0.1cm}
                    NW$_{45}$ & 0.09  & $5.46^{+0.07}_{-0.06}$ & $1.2^{+0.9}_{-0.8}$ & 9 & N.C.$^{(b)}$  & N.C.$^{(b)}$ & 4 & 0.05 & $2.84^{+0.1}_{-0.1}$ & $0.49^{+0.10}_{-0.08}$\\
                    \vspace{0.1cm}
                    N$_{012}$  & 0.08 & $5.46^{+0.04}_{-0.04}$ & $3^{+3}_{-2}$ & 8 & $5.94^{+0.05}_{-0.04}$ & $2.3^{+1.9}_{-1.7}$ & 5 & 0.07 & $2.76^{+0.18}_{-0.17}$ & $0.6^{+0.2}_{-0.2}$\\
                    \vspace{0.1cm}
                    N$_{123}$ & 0.12 & $5.46^{+0.04}_{-0.04}$  & $3^{+2}_{-2}$ & 7 & $5.92^{+0.04}_{-0.03}$ & $4^{+2}_{-2}$ & 11 & 0.09 & $2.72^{+0.16}_{-0.16}$ & $0.7^{+0.2}_{-0.2}$ \\
                    \vspace{0.1cm}
                    N$_{234}$ & 0.14 & $5.46^{+0.04}_{-0.03}$ & $3^{+3}_{-2}$ & 8 & $5.91^{+0.04}_{-0.03}$ & $4^{+2}_{-2}$ & 14 & 0.09 & $2.68^{+0.16}_{-0.16}$ & $0.6^{+0.2}_{-0.1}$\\
                    \vspace{0.1cm}
                    N$_{345}$ & 0.13 & N.C.$^{(b)}$ & N.C.$^{(b)}$ & 1 & $5.89^{+0.03}_{-0.03}$ & $4^{+2}_{-2}$ & 14 & 0.07 & $2.76^{+0.18}_{-0.18}$ & $0.54^{+0.02}_{-0.02}$ \\ 
                    \vspace{0.1cm}
                    NE$_{01}$  & 0.07 & $5.46^{+0.18}_{-0.07}$ & $1.3^{+1.0}_{-1.0}$ & 5 & N.C.$^{(b)}$ & N.C.$^{(b)}$ & 3 & 0.06 & $2.94^{+0.11}_{-0.11}$ & $0.7^{+0.1}_{-0.1}$\\
                    \vspace{0.1cm}
                    NE$_{12}$ & 0.08 & $5.38^{+0.21}_{-0.15}$ &  $1.2^{+1.0}_{-1.0}$ & 7 &  N.C.$^{(b)}$ &   N.C.$^{(b)}$ & 2 & 0.06 & $3.0^{+0.13}_{-0.12}$ & $0.6^{+0.2}_{-0.1}$\\
                    \vspace{0.1cm}
                    NE$_{23}$& 0.09 & $5.48^{+0.05}_{-0.05}$ & $2.1^{+1.1}_{-1.1}$ & 14 & $5.98^{+0.04}_{-0.04}$ & $1.9^{+1.0}_{-0.9}$ & 14 & 0.06 & $2.87^{+0.13}_{-0.12}$ & $0.48^{+0.11}_{-0.09}$\\
                    \vspace{0.1cm}
                    NE$_{34}$ & 0.10 & $5.49^{+0.03}_{-0.03}$ & $2.4^{+1.1}_{-1.0}$ & 18 & $5.94^{+0.04}_{-0.03}$ & $2.0^{+1.0}_{-0.9}$ & 15 & 0.07 & $2.77^{+0.12}_{-0.12}$ & $0.42^{+0.11}_{-0.08}$\\
                    \vspace{0.1cm}
                    NE$_{45}$ & 0.10 & $5.49^{+0.13}_{-0.37}$ & $<1.8$ & 3 & $5.92^{+0.05}_{-0.05}$ & $1.1^{+0.9}_{-0.8}$ & 5 & 0.07 & $2.81^{+0.14}_{-0.13}$ & $0.36^{+0.07}_{-0.08}$\\
                    \vspace{0.1cm}
                    C$_{NE}$ & 0.03 & $5.43^{+0.05}_{-0.05}$ & $1.1^{+0.9}_{-0.8}$ & 5 & $5.97^{+0.09}_{-0.12}$ & $0.72^{+0.77}_{-0.70}$ & 3 & 0.02 & $2.91^{+0.16}_{-0.16}$ & $0.27^{+0.09}_{-0.07}$ \\
                    \vspace{0.1cm}
                    C$_{NW}$ & 0.03 & N.C.$^{(b)}$ & N.C.$^{(b)}$ & 1 & $5.84^{+0.04}_{-0.06}$ & $<1.5$ & 8 & 0.03 & $3.00^{+0.18}_{-0.16}$ & $0.35^{+0.12}_{-0.09}$\\
                    \vspace{0.1cm}
                    C$_{SE}$ & 0.02 & N.C.$^{(b)}$  & N.C.$^{(b)}$  & 1 & $5.85^{+0.04}_{-0.04}$ & $1.2^{+0.7}_{-0.6}$ & 11 & 0.02 & $3.14^{+0.16}_{-0.16}$ & $0.40^{+0.12}_{-0.09}$\\
                    \vspace{0.1cm}
                    C$_{SW}$ & 0.02 & $5.52^{+0.05}_{-0.06}$ & $1.5^{+0.8}_{-0.7}$ & 13 & N.C.$^{(b)}$ & N.C.$^{(b)}$ & 2 & 0.02 & $2.76^{+0.18}_{-0.17}$ & $0.19^{+0.08}_{-0.05}$\\
                    \vspace{0.1cm}
                    S$_{012}$ & 0.02 & N.C.$^{(b)}$  & N.C.$^{(b)}$ & 0.5 & N.C.$^{(b)}$ & N.C.$^{(b)}$ & 1 & 0.05 & $2.87^{+0.18}_{-0.17}$ & $0.8^{+0.3}_{-0.2}$\\
                    \vspace{0.1cm}
                    S$_{123}$& 0.04 & $5.45^{+0.05}_{-0.05}$ & $3^{+2}_{-2}$ & 5 &  N.C.$^{(b)}$ &   N.C.$^{(b)}$ & 1 & 0.06 & $3.07^{+0.17}_{-0.16}$ & $1.2^{+0.4}_{0.3}$\\
                    \vspace{0.1cm}
                    S$_{234}$ & 0.05 & N.C.$^{(b)}$ &  N.C.$^{(b)}$ & 2 & N.C.$^{(b)}$  & N.C.$^{(b)}$ & 0 &0.06 & $3.03^{+0.16}_{-0.15}$ & $1.2^{+0.5}_{-0.3}$\\
                    \vspace{0.1cm}
                    S$_{345}$ & 0.05 & $5.43^{+0.06}_{-0.06}$&  $3^{+2}_{-2}$ & 5 & N.C.$^{(b)}$ &  N.C.$^{(b)}$ & 1 & 0.06 & $3.00^{+0.17}_{-0.16}$ & $1.0^{+0.3}_{-0.3}$\\
                        \hline
            \hline
                \end{tabular}
            \\$^{(a)}$ $\Delta$C$_{\rm{stat}}$ calculated as the difference of the  C$_{\rm{stat}}$ of the model without and with the line. \\ $^{(b)}$ Not Constrained (N.C.) or unreasonable values for lower and upper limits. \\ $^{(c)}$ At 1 keV.
                \label{tab:gauss}
\end{table*}

    \begin{figure}[h!]
    \centering
    \includegraphics[width=\textwidth]{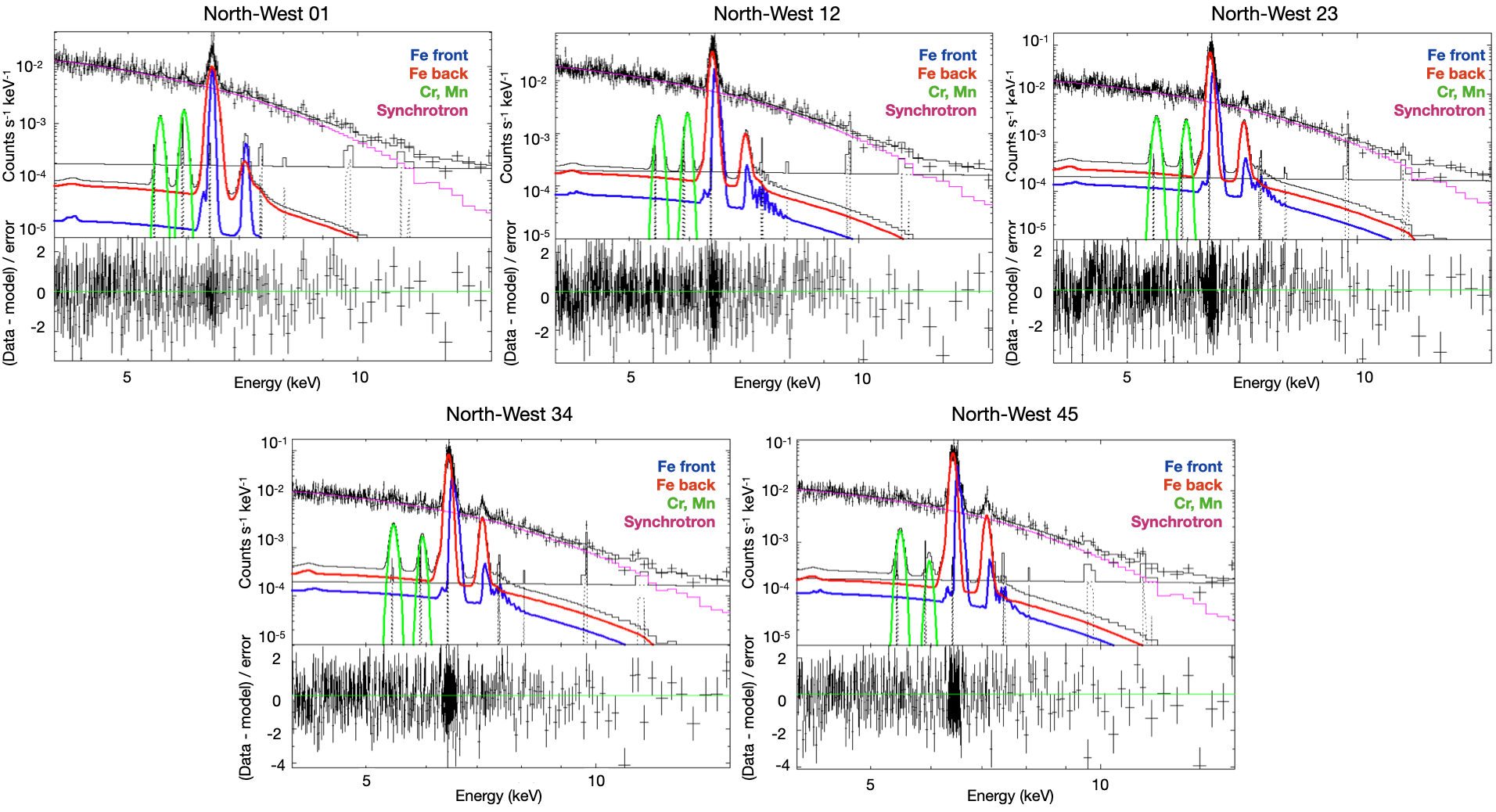}
    \caption{Spectra extracted from two rows of pixels (12 pixels in total) of the NW observation, fitted with the model in Eq. \ref{eq:mod_gauss}}
    \label{fig:spec_NW_rows}
    \end{figure}

    \begin{figure}[h!]
    \centering
    \includegraphics[width=\textwidth]{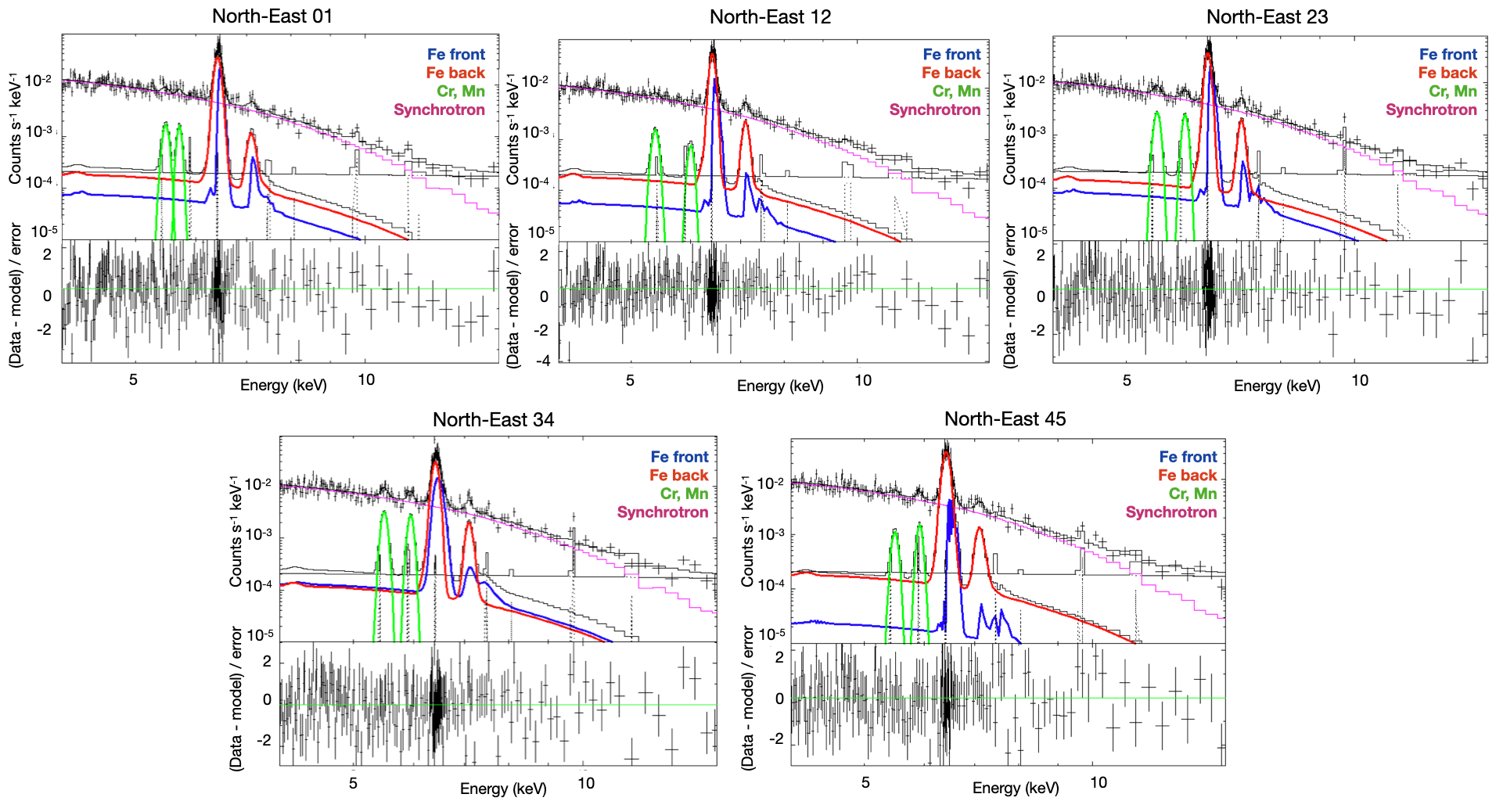}
    \caption{Spectra extracted from two rows of pixels (12 pixels in total) of the NE observation, fitted with the model in Eq. \ref{eq:mod_gauss}}
    \label{fig:spec_NE_rows}
    \end{figure}

    \begin{figure}[h!]
    \centering
    \includegraphics[scale=0.4]{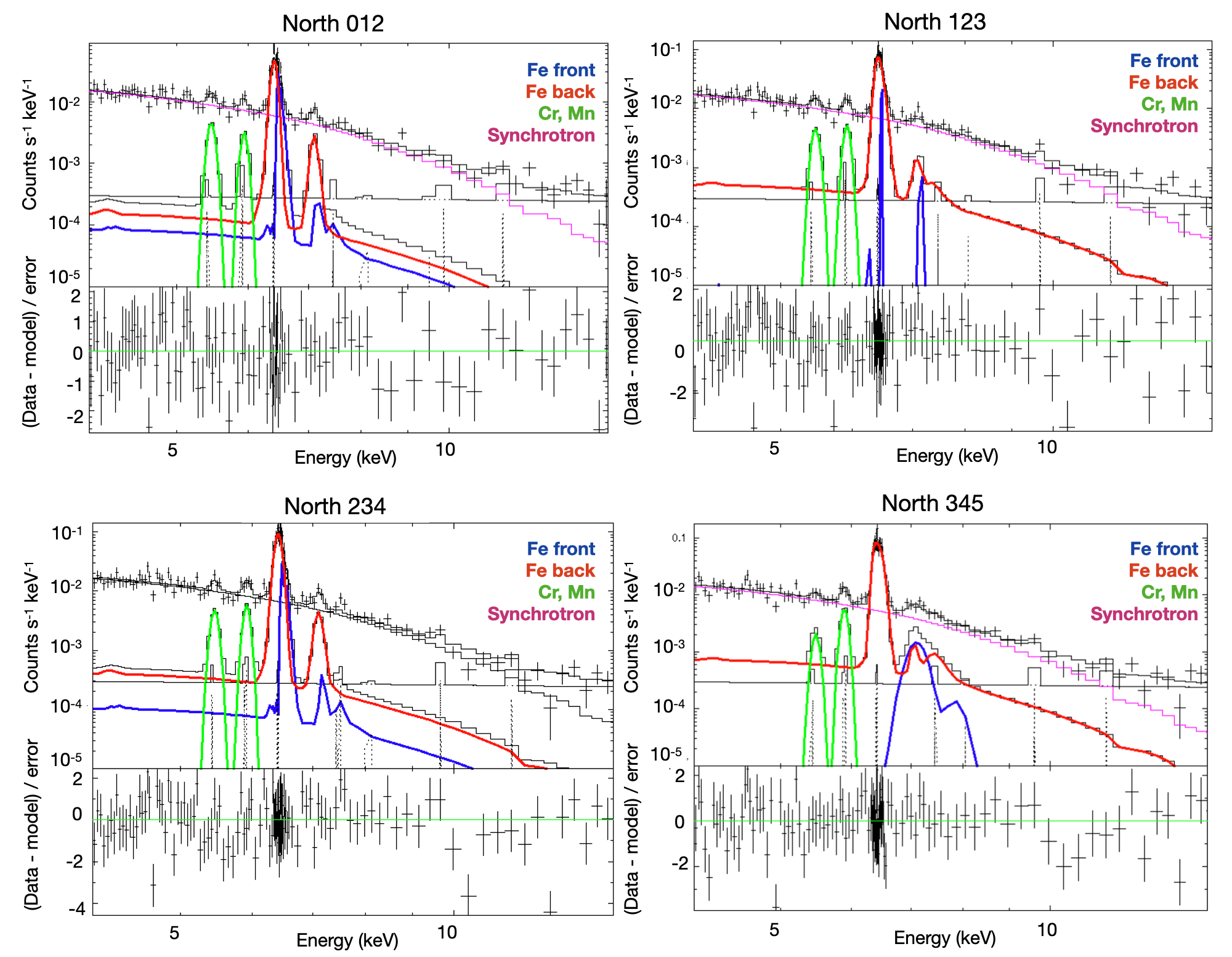}
    \caption{Spectra extracted from three rows of pixels (18 pixels in total) of the N observation, fitted with the model in Eq. \ref{eq:mod_gauss}}
    \label{fig:spec_N_rows}
    \end{figure}

    \begin{figure}[h!]
    \centering
    \includegraphics[scale=0.6]{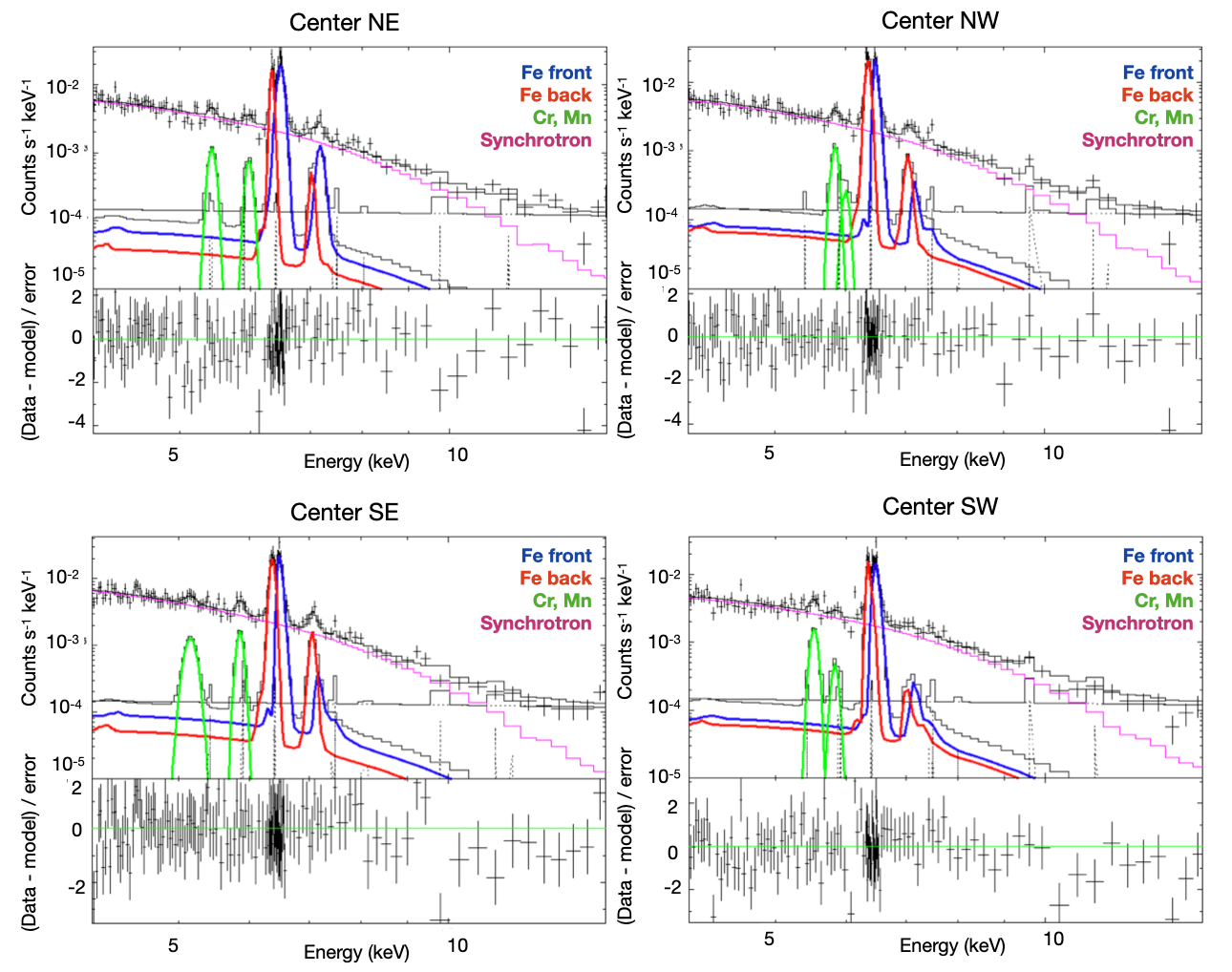}
    \caption{Spectra extracted from squares of pixels (9 pixels in total) of the C observation, fitted with the model in Eq. \ref{eq:mod_gauss}}
    \label{fig:spec_C_squares}
    \end{figure}

    \begin{figure}[h!]
    \centering
    \includegraphics[scale=0.6]{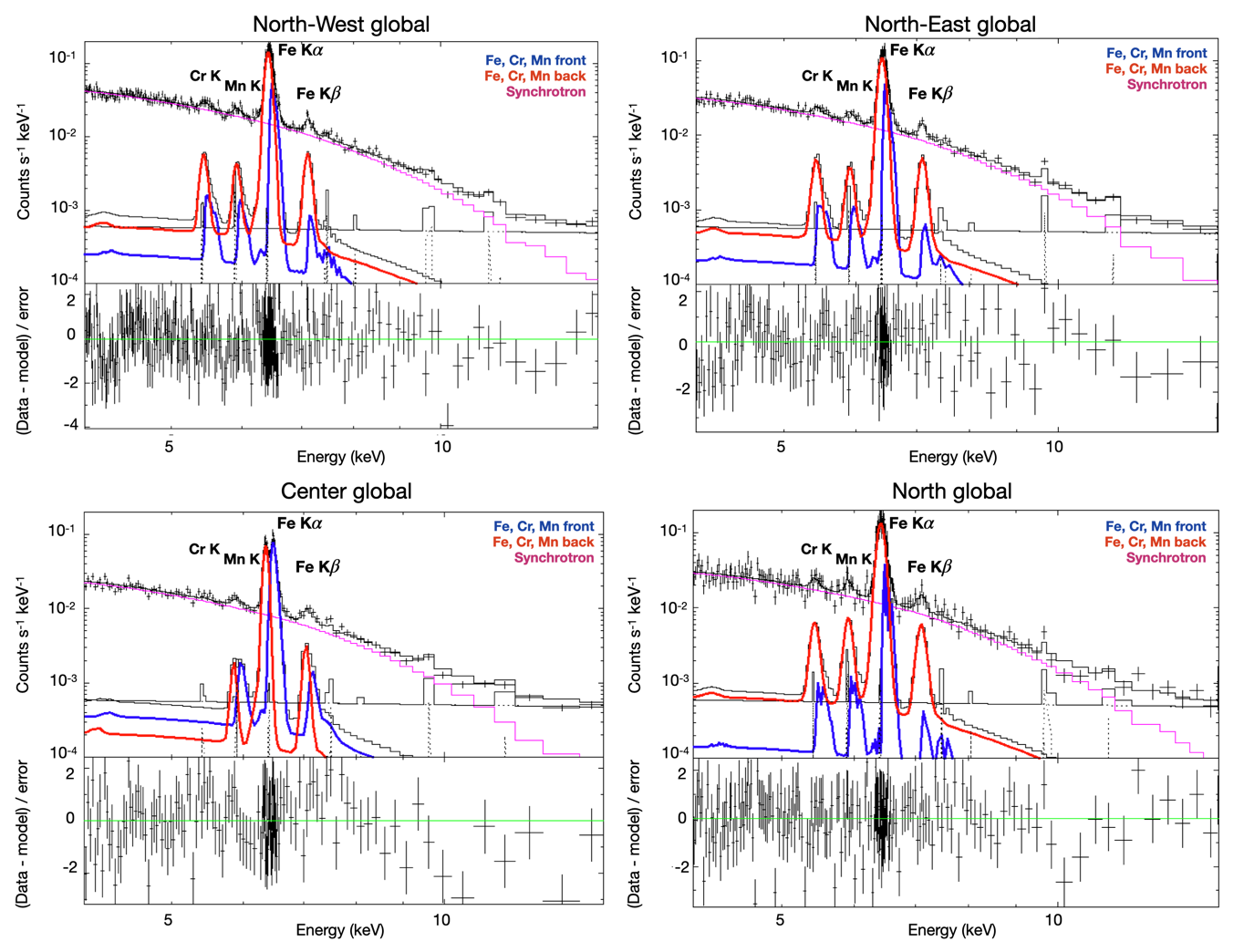}
    \caption{Spectra of the global pointings fitted with the model in Eq. \ref{eq:mod_pshock}}
    \label{fig:spec_global_pshock}
    \end{figure}

    \begin{figure}[h!]
    \centering
    \includegraphics[scale=0.4]{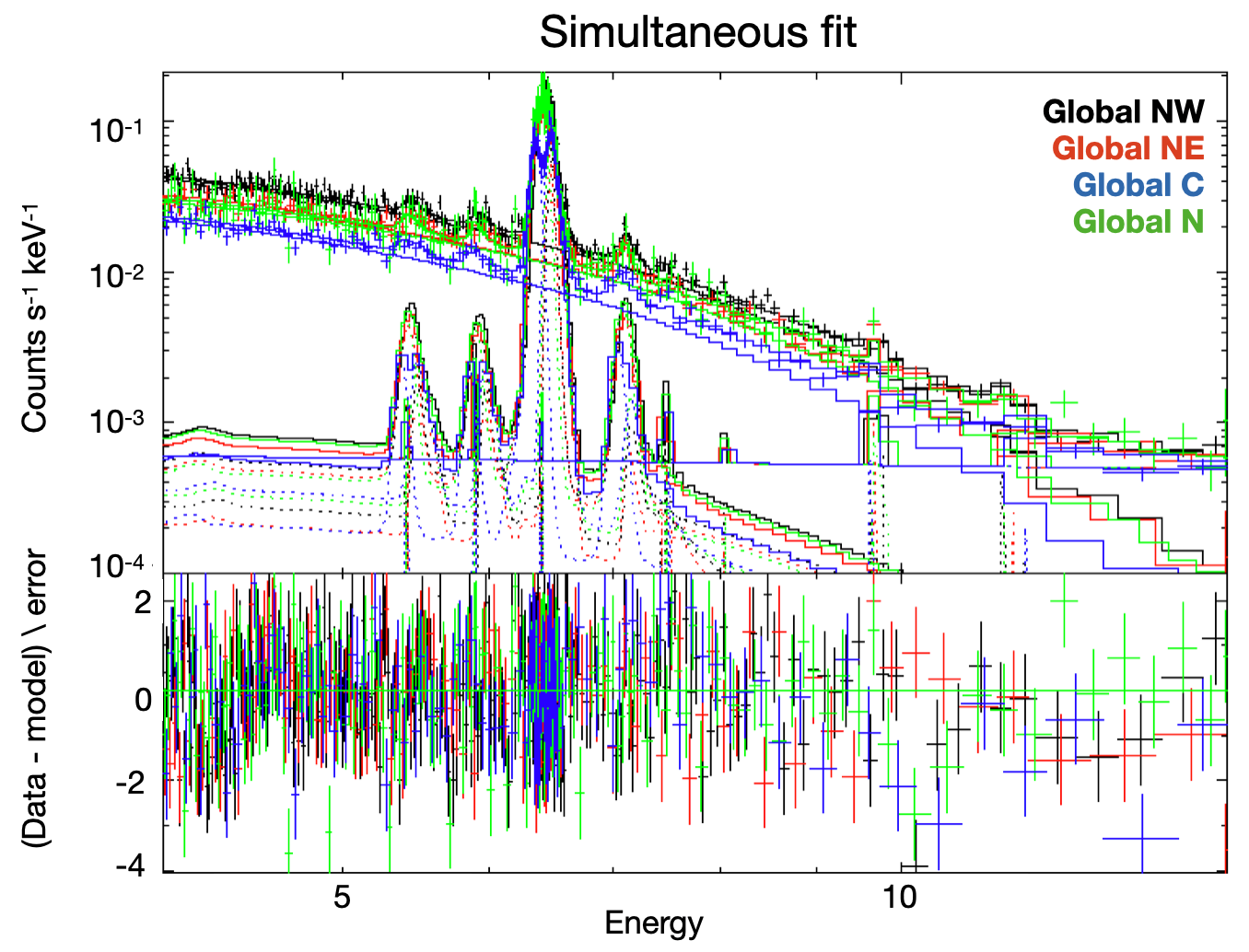}
    \caption{Simultaneous fit of all global spectra fitted with the model in Eq. \ref{eq:mod_pshock}}
    \label{fig:global_spec_average}
    \end{figure}

\end{onecolumn}
\end{document}